\documentclass[12pt]{article}
\usepackage{palatino}

\usepackage{amsfonts,amsmath,amsthm,amssymb,verbatim,nicefrac}
\usepackage{thmtools}

\usepackage[pdftex,letterpaper,left=0.95in,top=0.95in,bottom=0.95in,right=0.95in]{geometry} 
\usepackage{afterpage} 
\usepackage[round,longnamesfirst]{natbib}
\usepackage{graphicx, color}
\usepackage{setspace}
\usepackage[usenames,dvipsnames]{xcolor}
\usepackage[colorlinks,pagebackref=false]{hyperref}
\usepackage{epsf}
\usepackage{tikz}
\usetikzlibrary{shapes,snakes}
\usetikzlibrary{arrows.meta,calc,decorations.markings,math,arrows.meta}
\usepackage{pgfplots}
\pgfplotsset{compat=newest}
\usetikzlibrary{patterns,calc}
\usepackage{subcaption}
\usepackage{latexsym, stmaryrd, bm}
\usepackage[titletoc,title]{appendix}
\usepackage{titlesec}
\usepackage[dotinlabels]{titletoc}
\usepackage{marginnote}
\usepackage{ftnxtra}
\usepackage{todonotes}
\usepackage{minitoc}
\usepackage[nottoc]{tocbibind} 
\usepackage{enumitem}
\setlist{noitemsep,parsep=6pt,partopsep=0pt,topsep=0pt}

\definecolor{dark-red}{rgb}{0.60,0.15,0.15}
\definecolor{dark-blue}{rgb}{0.15,0.15,0.65}
\definecolor{light-blue}{rgb}{.2,1,1}
\definecolor{medium-blue}{rgb}{0,0,0.5}
\definecolor{dark-green}{rgb}{0,.5,0}

\hypersetup{
    pdftitle={Beyond Unbounded Beliefs},
    pdfauthor={Kartik, Lee, Liu, Rappoport},
    citecolor={dark-blue},
    bookmarksnumbered=true,
    urlcolor=dark-red,
    linkcolor=dark-red,
}

\declaretheorem{theorem}
\declaretheorem{lemma}
\declaretheorem{claim}
\declaretheorem{proposition}
\declaretheorem{conjecture}

\declaretheorem{corollary}

\declaretheorem[style=definition]{definition}
\declaretheorem[style=definition,qed=$\square$]{example}
\declaretheorem[style=definition]{condition}

\declaretheorem[style=remark]{remark}

  \newenvironment{thmbis}[1]
  {\renewcommand{\thetheorem}{\ref{#1}$'$}%
   \begin{theorem}}
  {\end{theorem}}

\newcommand{\citepos}[1]{\citeauthor{#1}'s (\citeyear{#1})}
\newcommand{\citeauthorpos}[1]{\citeauthor{#1}'s}

\makeatletter
  \renewcommand\@seccntformat[1]{\csname the#1\endcsname.{\hskip.7em\relax}} 
\makeatother

\titlespacing\section{0pt}{10pt plus 2pt minus 2pt}{4pt plus 2pt minus 2pt} 
\titlespacing\subsection{0pt}{6pt plus 2pt minus 2pt}{2pt plus 2pt minus 2pt} 
\titlespacing\subsubsection{0pt}{6pt plus 2pt minus 2pt}{0pt plus 2pt minus 2pt} 

\renewcommand{\epsilon}{\varepsilon}
\renewcommand{\phi}{\varphi}
\renewcommand{\bar}{\overline}
\newcommand{\mailto}[1]{\href{mailto:#1}{\texttt{#1}}}
\def\e{\epsilon}

\DeclareMathOperator*{\Supp}{Supp}

\DeclareMathOperator*{\argmax}{arg\,max}

\def\bi{\begin{itemize}}
\def\ei{\end{itemize}}

\def\E{\mathbb{E}}
\def\P{\mathbb{P}}
\def\Reals{\mathbb{R}}
\def\Integers{\mathbb{Z}}
\def\Naturals{\mathbb{N}}

\def\TP2{\text{TP}_2}
\newcommand\dif{\mathop{}\!\mathrm{d}}

\let\oldfootnote\footnote
\renewcommand\footnote[1]{\oldfootnote{\hspace{.4mm}#1}}

\makeatletter
\renewenvironment{proof}[1][\proofname] {\par\pushQED{\qed}\normalfont\topsep6\p@\@plus6\p@\relax\trivlist\item[\hskip\labelsep\bfseries#1\@addpunct{.}]\ignorespaces}{\popQED\endtrivlist\@endpefalse}
\makeatother

\let\oldFootnote\footnote
\newcommand\nextToken\relax

\renewcommand\footnote[1]{%
    \oldFootnote{#1}\futurelet\nextToken\isFootnote}

\newcommand\isFootnote{%
    \ifx\footnote\nextToken\textsuperscript{,}\fi}

\begin{document}

\pagestyle{empty}

\title{\color{BrickRed} Beyond Unbounded Beliefs:\\[3pt] \Large 
How Preferences and Information Interplay in Social Learning\footnote{We thank Nageeb Ali, Marina Halac, Ben Golub, Andreas Kleiner, Elliot Lipnowski, Jos\'{e} Montiel Olea, Xiaosheng Mu, Harry Pei, Jacopo Perego, Evan Sadler, Lones Smith, and Peter S{\o}rensen for helpful comments. We also received useful feedback from various seminar and conference audiences. C\'{e}sar Barilla, John Cremin, and Zikai Xu provided excellent research assistance. Kartik gratefully acknowledges support from NSF Grant SES-2018948.}}
\date{\today}
\author{Navin Kartik\thanks{Department of Economics, Columbia University. E-mail: \mailto{nkartik@columbia.edu}.} \and SangMok Lee\thanks{Department of Economics, Washington University in St. Louis. E-mail: \mailto{sangmoklee@wustl.edu}.} 
\and Tianhao Liu\thanks{Department of Economics, Columbia University. E-mail: \mailto{tl3014@columbia.edu}.}
\and Daniel Rappoport\thanks{Booth School of Business, University of Chicago. E-mail: \mailto{Daniel.Rappoport@chicagobooth.edu}.}}
\maketitle
\thispagestyle{empty}
\setstretch{1.2}



\begin{abstract}

When does society eventually learn the truth, or take the correct action, via observational learning?  In a general model of sequential learning over social networks, we identify a simple condition for learning dubbed \emph{excludability}. Excludability is a joint property of agents' preferences and their information. We develop two classes of preferences and information that jointly satisfy excludability: (i) for a one-dimensional state, preferences with single-crossing differences and a new informational condition, directionally unbounded beliefs; and (ii) for a multi-dimensional state, intermediate preferences and subexponential location-shift information. These applications exemplify that with multiple states ``unbounded beliefs'' is not only unnecessary for learning, but incompatible with familiar informational structures like normal information. Unbounded beliefs demands that a single agent can identify the correct action. Excludability, on the other hand, only requires that a single agent must be able to displace any wrong action, even if she cannot take the correct action.

\end{abstract}

\begin{quote}
\textbf{Keywords:} social learning; herds; information cascades; single crossing; Euclidean preferences; location-shift information; unbounded beliefs.
\end{quote}

\newpage
\setcounter{page}{1}
\pagestyle{plain}
\onehalfspacing

\section{Introduction}\label{sec:introduction}

This paper concerns the classic sequential observational or social learning model initiated by \citet{Banerjee92} and \citet{BHW92}. There is an unknown payoff-relevant state (e.g., product quality). Each of many agents has homogeneous preferences over her own action and the state (e.g., all prefer products of higher quality). Agents act in sequence, each receiving her own private information about the state and observing some subset of her predecessors' actions. The central economic question is about asymptotic learning: do Bayesian agents eventually learn to take the correct action (e.g., will the highest quality product eventually prevail)?

One would anticipate that whether there is social learning depends on the combination of agents' preferences and their information structure.  But, at least for finite action sets, economists have largely emphasized the latter dimension alone.\footnote{Unless noted otherwise, our introduction should be understood as referring to the canonical sequential social learning model with a finite action set, homogeneous preferences, and no direct payoff externalities. It is well recognized that variations in those aspects can also matter for social learning; see for example, \citet{Lee95} on infinite action spaces, \citet{AZ98} and \citet{EGKR14} on endogenous prices or congestion costs, and \citet{GPR06} on heterogeneous preferences.} The reason is inextricably tied to focusing on models with two states. With only two states, there is social learning given any (nontrivial) preferences if and only if there is learning for all preferences. For, with two states, even the former requires private signals/beliefs to be \emph{unbounded} \citep{SS2000,ADLO11}. Unbounded beliefs says that given any full-support prior it should be possible for a single private signal, however unlikely it is, to make an agent arbitrarily close to certain about the true state. 

With multiple---i.e., more than two---states, unbounded beliefs still characterizes learning for all preferences \citep{arieli2021general}.\footnote{\citet[Theorem 1]{arieli2021general} refer to the condition as ``totally unbounded beliefs''. They establish their result for a complete network, i.e., when each agent observes the actions of all predecessors. A by-product of our analysis is to establish it for general networks (\autoref{cor:UBsufficient} in \autoref{sec:main}).} However, it is now a very demanding condition. 
Consider, for instance, the canonical example of \emph{normal information}: the state is $\omega \in \Omega \subset \Reals$ and agents' signals are drawn independently from a normal distribution with mean $\omega$ and fixed variance. With only two states, there is unbounded beliefs because a very high signal makes one arbitrarily convinced of the high state, while a very low signal makes one arbitrarily convinced of the low state. But with multiple states, normal information fails unbounded beliefs: given any full-support prior, there is an upper bound on how certain one can become about any non-extremal state based on observing one signal.\footnote{So binary states is special because all states are extreme states. 
There is nothing exceptional about normal information violating unbounded beliefs; see \autoref{rem:MLRPnoUB} in \autoref{sec:main}. 
}  Is social learning doomed with multiple states for familiar information structures like normal information?

Our paper shows that the answer is no. With multiple states, whether society eventually learns to take the correct action
depends on the interplay of preferences and information. Crucially, learning can obtain under standard preferences with familiar information structures that fail unbounded beliefs. \autoref{fig:intro} illustrates an example of normal information 
with state space $\Omega=\{1,2,3\}$, action set $A=\{a_1,a_2\}$, and a uniform prior $\mu_0$.
The failure of unbounded beliefs is reflected in the set of posteriors, represented by the black curve, being bounded away from state 2's vertex. For concreteness, suppose that each agent observes all predecessors' actions. In \autoref{fig:intro2}, preferences violate single crossing---defined formally in \autoref{sec:applications}---because action $a_1$ is optimal in both states $1$ and $3$, whereas $a_2$ is optimal in state $2$. Here, learning fails: since action $a_1$ is optimal after any signal the first agent receives, society is stuck with all agents taking $a_1$. 
By contrast, in \autoref{fig:intro1}, agents have single-crossing preferences; specifically an agent who takes action $a_i$ gets the quadratic-loss utility $-(i-\omega)^2$. Now, at any belief at which learning the state would be useful (i.e., a belief that puts positive probability on both state $1$, where $a_1$ is optimal, and either state $2$ or $3$, where $a_2$ is optimal), no single action is optimal after all signals. 
This property yields social learning; see \autoref{lem:learning and stationary beliefs} in \autoref{sec:main}.

\begin{figure}[h] 
\centering 
    	\begin{subfigure}[b]{0.45\linewidth}
     \centering
		\begin{tikzpicture}[scale=1.2]
        \fill[blue!25] (2,0) -- (4,0) -- (3,1.732);
        \node[label={[blue] below:{\large $a_2$}}] at (3,1) {};

        \fill[blue!5] (0,0) -- (2,0) -- (3,1.732) -- (2,3.464);
        \node[label={[blue] below:{\large $a_1$}}] at (1.5,1.7) {};
  
		\coordinate (A) at (0,0); \coordinate (B) at (4,0); \coordinate (C) at (2,3.464); \coordinate (prior) at (2,1.2); \coordinate (prior23) at (2.97,1.78); \coordinate (prior12) at (2,0);
			\coordinate (D) at (2,0) ; \coordinate (E) at (1,1.5); \coordinate (F) at (3,1.5) ; 

		\draw[dashed] (A) node[below] {$1$} -- (B) node[below] {$2$} -- (C) node[above] {$3$} -- (A) ;
   
		\draw[thick] (A) to[out=0, in=300, looseness=1.6] (C);

		\node[circle,fill=black,inner sep=0pt,minimum size=3pt,label={below:{\large $\mu_0$}}] at (prior) {};
		\end{tikzpicture}
		\caption{$u(a_1,\omega)=-\frac{1}{2}$, $u(a_2,\omega)=-(2-\omega)^2$} \label{fig:intro2}
    \end{subfigure}
    \quad
	\begin{subfigure}[b]{0.45\linewidth}
 \centering
		\begin{tikzpicture}[scale=1.2]

        \fill[blue!5] (0,0) -- (2,0) -- (1,1.732);
        \node[label={[blue] right:{\large $a_1$}}] at (0.3,0.45) {};


        \fill[blue!25] (0.5,0.868) -- (2,0) -- (4,0) -- (2,3.464);
        \node[label={[blue] below:{\large $a_2$}}] at (3,1) {};
  
		\coordinate (A) at (0,0); \coordinate (B) at (4,0); \coordinate (C) at (2,3.464); \coordinate (prior) at (2,1.2); \coordinate (prior23) at (2.97,1.78); \coordinate (prior12) at (2,0);
			\coordinate (D) at (2,0) ; \coordinate (E) at (1,1.5); \coordinate (F) at (3,1.5) ; 

		\draw[dashed] (A) node[below] {$1$} -- (B) node[below] {$2$} -- (C) node[above] {$3$} -- (A) ;
   
		\draw[thick] (A) to[out=0, in=300, looseness=1.6] (C);

		\node[circle,fill=black,inner sep=0pt,minimum size=3pt,label={below:{\large $\mu_0$}}] at (prior) {};
        
		\end{tikzpicture}
  \caption{$u(a_i,\omega)=-(i-\omega)^2$}
  \label{fig:intro1}
    \end{subfigure}
    
\caption{Belief simplex for state space $\Omega=\{1,2,3\}$. The curve depicts the set of posteriors for a single agent under normal information with prior $\mu_0$. The action set is $A=\{a_1,a_2\}$ and each agent's utility is $u(a,\omega)$. The shaded regions depict optimal actions under uncertainty.}
\label{fig:intro}
\end{figure}

\paragraph{Excludability.} Our paper develops a simple joint condition on information and preferences, which we call \hyperref[def:exclude]{\emph{excludability}}, that is not only sufficient for social learning on general observational networks (satisfying a mild condition known as \hyperref[eq:exp obs]{expanding observations}), but in a sense also necessary; see \autoref{thm:exclude} in \autoref{sec:main}.

Roughly speaking, excludability requires that for each pair of actions, $a$ and $a'$, a single agent must be able to receive a signal that makes her arbitrarily convinced that $a$ is better than $a'$, no matter which (full-support) belief she starts with. Put differently, information must be able to distinguish the set of states in which $a$ is better than $a'$ from the set in which $a'$ is better than $a$. 
Excludability implies that society can never get stuck on a wrong action: if an action is suboptimal at the true state, then some agent will receive a private signal convincing her not to take that action. 
We establish that this property of \emph{displacing wrong actions} leads to social learning. 
Notably, an agent can displace wrong actions even if she cannot take the correct action, i.e., the optimal action at the true state.
(See \autoref{fig:simulation} in \autoref{sec:main} for a concrete example.)
We view the distinction of social learning arising from the individual capacity to displace wrong actions rather than to discover the correct action as a key insight; this distinction cannot be seen with only two states, where the two notions are equivalent.

Excludability provides a useful perspective on existing ideas in the literature. For instance, as detailed in \autoref{sec:main}, an information structure yields excludability for all preferences if and only if that information structure has unbounded beliefs. 
But more importantly, we can use excludability to deduce 
weaker informational conditions that yield social learning 
for canonical classes of preferences.\footnote{Although this approach of obtaining more tenable conditions by restricting preferences to some broad class is novel to social learning, it is classical in other areas of economics. For instance, first-order stochastic dominance is weakened to second-order by restricting to concave (and increasing) utility functions.} 

\paragraph{Single-crossing preferences.} Our leading application of excludability is to preferences with \emph{\hyperref[def:SCD]{single-crossing differences}} (SCD). Here we show that learning obtains when the information structure satisfies 
\emph{\hyperref[def:DUB]{directionally unbounded beliefs}} (DUB). SCD is a familiar property \citep{milgrom1994monotone} that is widely assumed in economics: it captures settings in which there are no preference reversals as the state increases. 
By contrast, DUB appears to be a new condition on information structures, although \citet{Milgrom79} utilizes a related property in the context of auction theory. Like SCD, DUB is formulated for a (totally) ordered state space. It requires that for any state $\omega$ and any prior that puts positive probability on $\omega$, there exist both: (i) 
signals that make one arbitrarily certain that the state is at least $\omega$; and (ii)  
signals that make one arbitrarily certain that the state is at most $\omega$. 
Crucially, no signal need make one arbitrarily certain about $\omega$ (unlike unbounded beliefs). For the normal information structure discussed earlier, requirements (i) and (ii) are met for any state by arbitrarily high and arbitrarily low signals, respectively.  

\autoref{thm:main} in \autoref{sec:applications} shows that SCD preferences and DUB information are jointly sufficient for excludability, and hence learning. For a direct intuition on 
the SCD-DUB interplay, consider normal information again. 
There are preferences (like those in \autoref{fig:intro2}) 
under which society can get stuck at some belief at which agents are taking an incorrect action, but only a strong signal about an intermediate state would change the action---alas no such signal is available. However, under SCD preferences (like those in \autoref{fig:intro1}), if knowing that the state is some intermediate $\omega$ would change the action, then so would knowing that the state is at least $\omega$ or at most $\omega$. Normal information, or more generally DUB, guarantees that there are strong signals approximating such knowledge.

\paragraph{Intermediate preferences.} 
Our second application in \autoref{sec:applications} 
is to \emph{\hyperref[def:HS]{intermediate preferences}} in multidimensional spaces \citep{Grandmont78}, where the state is $\omega\in \Reals^d$ and the action is $a\in \Reals^d$. These subsume both constant-elasticity-of-substitution preferences common in many areas of economics and Euclidean preferences invoked in political economy and communication/delegation models. 

Using excludability, we show that social learning obtains under intermediate preferences so long as information is given by a \emph{\hyperref[def:SLS]{subexponential location-shift family}}. Location-shift families are widely-used information structures: for some density $g: \Reals^d \to \Reals^d$, the signal distribution in any state $\omega$ is given by $g(s-\omega)$. Loosely, the subexponential condition requires that the tail of $g$ must be thin enough, eventually decreasing faster than an exponential rate. We establish that this thin-tails property combined with intermediate preferences yields excludability. Notably, multidimensional normal information (i.e., normally distributed signals with mean equal to the state and some fixed covariance matrix) 
satisfies the subexponential requirement.

\paragraph{Methodology.} 
A significant contribution of our paper is also methodological. We develop an approach to tackle learning,
and more generally, asymptotic social welfare with multiple states in general observational networks. \autoref{lem:learning and stationary beliefs} in \autoref{sec:main} is the backbone by which we tie learning to excludability. \autoref{lem:learning and stationary beliefs} reduces the complex dynamic problem of social learning in networks to a much simpler ``static'' problem. The theorem says that there is learning if and only if every \emph{stationary belief} has \emph{adequate knowledge}. A stationary belief is one at which there is an action that is optimal no matter an agent's signal, 
and an adequate-knowledge belief is one at which there is an action that is optimal no matter the state in the belief's support. 
Excludability is a simple sufficient---and necessary, in a sense explained later---condition for all stationary beliefs to have adequate knowledge. 

\autoref{lem:learning and stationary beliefs} itself is a consequence of
\autoref{thm:utility diffusion} in \autoref{sec:cascade utility}, which 
provides a welfare lower bound even when learning fails. The theorem roughly says that for any preferences and information (and given expanding observations), agents eventually obtain at least their \emph{cascade utility}. Cascade utility is the minimum expected utility an agent can get from any Bayes-plausible distribution of stationary beliefs. \autoref{thm:utility diffusion} implies that learning obtains when the cascade utility equals the utility obtained from taking the correct action in each state, which leads to \autoref{lem:learning and stationary beliefs}.

\paragraph{Related literature.}

A number of papers on sequential Bayesian social learning only consider the complete observational network: each agent observes all her predecessors' actions. For that case and with binary states, \citet{SS2000} 
show that, given any nontrivial preferences, there is learning if and only if beliefs are unbounded. For the complete network but with multiple states, \citet{arieli2021general} show that unbounded beliefs---which they call ``totally unbounded beliefs''---is sufficient for learning, and also necessary if learning must obtain no matter society's preferences.\footnote{The early work of \citet*{BHW92} allowed for multiple states, but  they only identified failures of learning because they implicitly restricted attention to ``bounded beliefs''; more precisely, they assumed finite signals with full-support distributions.}   The  approach of both \citet{SS2000} and \citet{arieli2021general} rests on 
the social belief---an agent's belief based on observing her predecessors' actions, before observing her own signal---being a martingale in the complete network.

\citet{GK03} and \citet{CK04} depart from the complete network, noting that martingale methods now fail. Both these papers also depart from the canonical setting in other ways, however: in \citet{GK03} agents choose actions repeatedly, while in \citet{CK04} private signals are not independent conditional on the true state. \citet*{ADLO11} provide a general treatment of observational networks in an otherwise classical setting. But they only allow for binary states and binary actions. 
They introduce the condition of expanding observations, explaining that this property of the network is necessary for learning. They 
establish that it is also sufficient for learning with unbounded beliefs. Building on \citet{BF04}, a key contribution of \citet*{ADLO11} is to use a welfare \emph{improvement principle} to deduce learning; 
this approach works even though martingale arguments fail. \citet{LS15} introduce a notion of ``information diffusion'' and use the improvement principle to establish information diffusion even when learning fails. 

The analysis in both \citet*{ADLO11} and \citet{LS15} relies on their binary-state binary-action structure.\footnote{\citet{BF04} and \citet{SS2020} consider ``unordered'' random sampling models that also only allow for binary states and actions.} We believe ours is the first paper to consider the canonical sequential social learning problem with general observational networks and general state and action spaces. At a methodological level, we develop a novel analysis based on continuity and compactness---rather than monotonicity or other properties that are specific to binary states or actions---that uncovers the fundamental logic underlying a general improvement principle. 

Substantively, our focus on multiple states and actions allows us to shed light on how preferences and information jointly shape social learning. As already noted, their interplay in determining learning has not received attention in the prior literature because of its focus on binary states. The only exception we are aware of is \citet[Theorem 3]{arieli2021general}, discussed in \autoref{sec:main}; their result assumes a special utility function and is only for the complete network.

\section{Model}
\label{sec:model}

There is a countable state space $\Omega$, endowed with the discrete topology, and standard Borel spaces of actions $A$ and signals $S$. We allow each of these three sets to be finite or infinite. An information or signal structure is given by a collection of probability measures over $S$, one for each state, denoted by $F(\cdot | \omega)$. Assume that for any $\omega$ and $\omega'$, $F(\cdot | \omega)$ and $F(\cdot | \omega')$ are mutually absolutely continuous. It follows that each $F(\cdot|\omega)$ has a density $f(\cdot|\omega)$; more precisely, this is the Radon-Nikodym derivative of $F(\cdot|\omega)$ with respect to some reference measure that is mutually absolutely continuous with every $F(\cdot|\omega')$. Without further loss of generality we assume $f(\cdot | \cdot) > 0$, so that no signal rules out any state.

\paragraph{The game.} At the outset, a state $\omega$ is drawn from a common prior probability mass function 
$\mu_0 \in \Delta \Omega$.\footnote{For any topological space $X$, $\Delta X$ denotes the set of Borel probability measures over $X$.
} Then, an infinite sequence of agents, indexed by $n=1,2,\ldots$, 
sequentially select actions. An agent $n$ observes both a private signal $s_n$ drawn 
from $f(\cdot|\omega)$ and the actions of some subset of her predecessors $B_n \subseteq \{1, 2, \dots, n-1\}$, and then chooses her action $a_n\in A$. Agents' private signals are drawn independently conditional on the state, and no agent observes either the state or any of her predecessors' signals. Each observational neighborhood $B_n$ is stochastically generated according to a probability distribution $Q_n$ over all subsets of $\{1, 2, \dots, n-1\}$, assumed to be independent across $n$, independent of the state $\omega$, and independent of any private signals. The distributions $(Q_n)_{n \in \mathbb N}$ constitute the observational network structure and are common knowledge, but the realized neighborhood $B_n$ is the private information of agent $n$.

Agent $n$'s information set thus consists of her signal $s_n$, neighborhood $B_n$, and the actions chosen by the neighbors $(a_k)_{k \in B_n}$.\footnote{While we assume that each agent observes the identities of her neighbors as well as their chosen actions, the \hyperref[sec:conclusion]{Conclusion} explains how our analysis extends to
various cases of ``random sampling'' in which neighbors' identities are not observed. Our analysis also applies if agents receive arbitrary information about their predecessors' realized neighborhoods.
}
Let $\mathcal I_n$ denote the set of all possible information sets for agent $n$.
A strategy for agent $n$ is a (measurable) function $\sigma_n: \mathcal I_n \to \Delta A$.

All agents are expected utility maximizers and have common preferences that depend only on their own action and the state, represented by the utility function $u:A \times \Omega \to \Reals$. We assume that utility is bounded: there is $\bar u \geq 0$ such that $|u(\cdot,\cdot)|\leq \bar u$.

We study the Bayes Nash equilibria---hereafter simply equilibria---of this game. 
We assume that for every belief there is an optimal action, so that an equilibrium exists.\footnote{\label{fn:BNE}
Existence of optimal actions is assured under standard assumptions, e.g., 
if $A$ is compact and $u(\cdot,\cdot)$ is suitably continuous.
We also note that as there are no direct payoff externalities, strategic interaction is minimal: any $\sigma_n$ affects other agents only insofar as affecting how $n$'s successors update about signal $s_n$ from the observation of action $a_n$. Hence, we could just as well adopt (weak) Perfect Bayesian equilibrium or refinements.}

\begin{remark}\label{rmk:discontinuity}
\autoref{app:backbone} describes a more general setting in which our main results are proved. 
For example, $\Omega$ can be a closed subset of $\Reals$ and each $u(a,\cdot)$ piecewise continuous with $A$ finite. We also do not require the signal distributions to be mutually absolutely continuous.
\end{remark}

\paragraph{Adequate learning.} 

The full-information expected utility given a belief $\mu$ is the expected utility under that belief if the state will be revealed before an action is chosen: 
$$
u^*(\mu) := \sum_{\omega\in \Omega} \max_{a\in A} u(a, \omega) \mu(\omega). 
$$
Given a prior $\mu_0$ and a strategy profile $\sigma$, agent $n$'s utility $u_n$ is a random variable. Let $\E_{\sigma, \mu_0} [u_n]$ be agent $n$'s ex-ante expected utility. We say there is \emph{adequate learning} if for every prior $\mu_0$ and every equilibrium $\sigma$, 
$\E_{\sigma, \mu_0} [u_n] \to u^*(\mu_0)$. In words, adequate learning requires that given any prior and equilibrium, no matter which state is realized, eventually agents take actions that are arbitrarily close to optimal in that state.\footnote{
Our notion of adequate learning is different from \citepos{arieli2021general}, who require learning for all utility functions. Following \citet{ABHJ91}, we use ``adequate'' to signify that learning the state precisely is not necessary when some action is optimal in multiple states.}
We say there is \emph{inadequate learning} if adequate learning fails.\footnote{That we deem learning to be inadequate if there is some equilibrium in which learning fails, rather than in every equilibrium, is innocuous given that there is no strategic interaction (cf.~\autoref{fn:BNE}). On the other hand, the issue of whether learning fails at every prior rather than only at some priors is substantive. 
We return to this issue in our \hyperref[sec:conclusion]{Conclusion}.
}

We will also be interested in situations in which agents choose from a subset of actions, referred to as a \emph{choice set}.\footnote{We restrict attention to choice sets such that for every belief there is an optimal action.}
We say that there is \emph{(in)adequate learning for a choice set $\tilde A\subseteq A$} if there is (in)adequate learning when agents are restricted to choose from actions in $\tilde A$. 

\paragraph{Expanding observations.}
As observed by \citet*{ADLO11}, a necessary condition for adequate learning is that the network structure has \emph{expanding observations}: 
\begin{equation}\label{eq:exp obs}
    \forall K \in \mathbb N: \ \lim_{n \to \infty} Q_n\left(
B_n \subseteq \{1, \dots, K\}\right) = 0.
\end{equation}
The reason is that a failure of expanding observations means that for some $K\in \mathbb{N}$, there is an infinite number of agents each of whom, with probability uniformly bounded away from $0$, observes at most actions $a_1,\ldots,a_K$. In that event, the agent cannot do better than choosing her action based on only $K+1$ signals.

Accordingly, we assume expanding observations. Leading examples of network structures with expanding observations include: (i) the classic complete network in which each agent's neighborhood is all her predecessors (formally, $Q_n(B_n=\{1,\ldots,n-1\})=1$); (ii) each agent only observes her immediate predecessor ($Q_n(B_n=\{n-1\})=1$); and (iii) each agent observes a random predecessor ($Q_n(B_n=\{k\})=1/(n-1)$ for all $k\in \{1,\ldots,n-1\}$).

\section{Characterizations of Learning}
\label{sec:main}

\subsection{Stationary Beliefs and Adequate Knowledge}

The key to all our results on learning is \autoref{lem:learning and stationary beliefs} below, which simplifies the question of adequate learning to a ``one-shot updating'' property of beliefs. To state that result, we require two concepts concerning the value of information.

For any belief $\mu\in \Delta \Omega$, let $c(\mu):=\argmax_{a\in A} \E_\mu [u(a, \omega)]$ denote the set of optimal actions under that belief.
Abusing notation, for a degenerate belief on state $\omega$ we write $c(\omega)$. Denoting the posterior after signal $s$ when starting from belief $\mu$ by $\mu_s$, we say that belief $\mu$ is \emph{stationary} if there is $a\in c(\mu)$ such that $a\in c(\mu_s)$ for $\mu$-a.e.~signal $s$. We say that belief $\mu$ has \emph{adequate knowledge} if there is $a\in c(\mu)$ such that $a\in c(\omega)$ for all $\omega\in \Supp \mu$.
So a belief is stationary if an agent holding that belief does not benefit from observing a signal from the given information structure.\footnote{Some readers may find it helpful to note that in their setting, \citet{SS2000} refer to stationary beliefs as ``cascade beliefs''.} On the other hand, a belief has adequate knowledge if the agent would not benefit from observing a signal from \emph{any} information structure, in particular learning the state.

Any adequate-knowledge belief, such as a belief that puts probability one on a single state, is stationary. In general, there can be stationary beliefs without adequate knowledge, as seen in \autoref{fig:intro2}. 

\begin{theorem}\label{lem:learning and stationary beliefs}
There is adequate learning if and only if all stationary beliefs have adequate knowledge.
\end{theorem}

\autoref{lem:learning and stationary beliefs} provides a characterization of adequate learning that holds regardless of the observational network structure, given our maintained assumption of expanding observations. Its ``only if'' direction is straightforward because our notion of learning considers all priors: if the prior is stationary and has inadequate knowledge, then society is stuck with all agents taking the prior-optimal action even though it is suboptimal in some states. More important and subtle is the theorem's ``if'' direction. It is inspired by earlier results, particularly  \citet[Lemma 1]{arieli2021general} and \citet[Theorem 1]{LS15}, but the logic in the current general setting of arbitrary networks and multiple states and actions is novel. 
We defer this logic to \autoref{sec:cascade utility}, instead turning now to how we can build on \autoref{lem:learning and stationary beliefs} for a more practicable characterization of learning. In particular, we seek a more transparent condition on the combinations of preferences and information that yield adequate learning.

\subsection{Excludability}


A key notion is whether information allows an agent to become arbitrarily sure about a subset of states $\Omega'$ relative to another 
subset $\Omega''$. 
To make that precise, let $\mu_s(\Omega')$ denote the posterior on states $\Omega'$ induced by belief $\mu$ and signal $s$, and $\Pr_\mu(S')$ be the probability of signal set $S'$ induced by belief $\mu$.

\begin{definition}\label{def:distinguishability}
A set $\Omega'$ is \emph{distinguishable} from another set $\Omega''$ if for any $\epsilon>0$ and \mbox{$\mu\in \Delta (\Omega'\cup\Omega'')$} with $\mu(\Omega')>0$, 
it holds that $\Pr_\mu(s:\mu_s(\Omega')>1-\epsilon)>0$.
\end{definition}

Note that $\Omega'$ is distinguishable from $\Omega''$ if and only if every $\omega\in \Omega'$ is distinguishable from $\Omega''$. Moreover, if $\Omega'$ is distinguishable from $\Omega''$, then every subset of $\Omega'$ is distinguishable from every subset of $\Omega''$. The following observation essentially reinterprets distinguishability directly in terms of the signal structure rather than 
posteriors.

\begin{lemma}
\label{lem:DUB-Distinguishability}
$\Omega'$ is distinguishable from $\Omega''$ if for every $\omega'\in \Omega'$ and $\epsilon > 0$, there is a positive-probability set of signals $S'$ such that
$$\forall \omega''\in \Omega'',  \forall s \in S': 
\frac{f(s|\omega'')}{f(s|\omega')} < \epsilon.$$
Conversely, this condition is also necessary if $\Omega''$ is finite.
\end{lemma}


We emphasize that the set $S'$ in the lemma cannot depend on $\omega''\in \Omega''$; for $\Omega'$ to be distinguished from $\Omega''$, each $\omega'\in \Omega'$ must be distinguished from all $\omega''\in \Omega''$ {simultaneously}. Consider the example of \emph{normal information}: $\Omega \subset \Reals$ and signals are normally distributed on $\Reals$ with mean $\omega$ and fixed variance. When $\Omega=\{1,2,3\}$, state $2$ is distinguishable from $1$ because $f(s|1)/f(s|2)\to 0$ as $s\to \infty$, and state $2$ is distinguishable from $3$ because $f(s|3)/f(s|2)\to 0$ as $s\to -\infty$. But state $2$ cannot be distinguished from both $1$ and $3$ simultaneously, because $\min\{f(s|1)/f(s|2),f(s|3)/f(s|2)\}$ is bounded away from $0$.


Distinguishability of each state from its complement is the condition of \emph{unbounded beliefs}; this is termed ``totally unbounded beliefs'' by \citet{arieli2021general} and is the multi-state extension of the two-state notion introduced by \citet{SS2000}. But with multiple states, unbounded beliefs is incompatible with familiar information structures. 
\begin{remark}
\label{rem:MLRPnoUB}Under any monotone likelihood ratio property (MLRP) information structure, no state $\omega$ is distinguishable from $\{\omega',\omega''\}$ with $\omega'<\omega<\omega''$.\footnote{\label{fn:MLRPdef}For ordered state and signals spaces, the MLRP holds if $\forall s'>s$ and $\forall \omega'>\omega$,  ${f(s|\omega')}/{f(s|\omega)} \leq {f(s'|\omega')}/{f(s'|\omega)}$.  } Consequently, 
if $|\Omega|>2$, unbounded beliefs fails under the MLRP.
\end{remark}

Fortunately, learning only requires certain subsets of states to be distinguished from each other. For any two actions $a_1$ and $a_2$, let the preferred set  $\Omega_{a_1,a_2} := \{\omega: u(a_1,\omega) >u(a_2,\omega)\}$ 
be the set of states in which $a_1$ is strictly preferred to $a_2$.

\begin{definition}
\label{def:exclude}
 A utility function and an information structure jointly satisfy \emph{excludability} if for every 
 $a_1$ and $a_2$, $\Omega_{a_1,a_2}$ is distinguishable from $\Omega_{a_2,a_1}$.
\end{definition}

Excludability is a joint condition on 
preferences and information. It requires that for any pair of actions, a single agent can become arbitrarily certain that one action is strictly better than the other, starting from any belief that does not exclude that event. Since excludability is defined using preferred sets, it is straightforward to deduce which sets must be distinguishable for any given preferences; \autoref{lem:DUB-Distinguishability}  then provides a set of likelihood-ratio conditions on the information structure, without reference to beliefs.



Unbounded beliefs implies excludability for any preferences. Conversely, if unbounded beliefs fails, then there is some state $\omega^*$ that is not distinguishable from its complement, and excludability fails when preferences are such that for some $a_1$ and $a_2$, $\Omega_{a_1,a_2}=\{\omega^*\}$ while $\Omega_{a_2,a_1}=\Omega\setminus\{\omega^*\}$. Hence, excludability for all preferences is equivalent to unbounded beliefs. But with multiple states, excludability can be substantially weaker for any given (class of) preferences, as developed in \autoref{sec:applications}.\footnote{With only two states, $\Omega=\{\omega_1,\omega_2\}$, excludability under any given nontrivial preferences is equivalent to unbounded beliefs. (Nontrivial means that no action is optimal at all states.) For, there must be actions $a_1$ and $a_2$ such that $\Omega_{a_1,a_2}=\{\omega_1\}$ and $\Omega_{a_2,a_1}=\{\omega_2\}$; excludability requires these sets to be mutually distinguishable, which is unbounded beliefs.}
 This matters because:

\begin{theorem}\label{thm:exclude}
Excludability implies adequate learning for every choice set. If excludability fails and the number of states is finite, then there is inadequate learning for some choice set. 
\end{theorem}

(See \autoref{exclude_prime} in the appendix for a more general version of \autoref{thm:exclude} that does not require finiteness in the second statement. Hereafter, for brevity, we leave it as implicit that it is \autoref{exclude_prime} rather than \autoref{thm:exclude} we are invoking when discussing the necessity of excludability for learning in an infinite state space.)

Excludability is sufficient for adequate learning because it ensures that wrong actions can always be ``displaced'', which by \autoref{lem:learning and stationary beliefs} is the key to social learning. More precisely, excludability guarantees that, no matter the choice set, all stationary beliefs have adequate knowledge. Suppose a belief $\mu$ has \emph{in}adequate knowledge, so that $c(\mu) \neq c(\omega^*)$ for some state $\omega^* \in \Supp \mu$. (For simplicity, assume $c(\mu)$ and $c(\omega^*)$ are singletons.) Excludability implies that preferred set $\Omega_{c(\omega^*),c(\mu)}$ is distinguishable from $\Omega_{c(\mu),c(\omega^*)}$. Hence, with positive probability, an agent who starts with belief $\mu$ will obtain a posterior that puts arbitrarily large probability on $\Omega_{c(\omega^*),c(\mu)}$ relative to $\Omega_{c(\mu),c(\omega^*)}$, in which event she strictly prefers $c(\omega^*)$ to $c(\mu)$. Consequently, $\mu$ is not stationary.

We highlight that excludability does not guarantee that a wrong action can always be displaced by the correct action. In other words, even though excludability guarantees that given any wrong action---say, $c(\mu)$ when the true state is $\omega^*$---a single agent can receive a signal  convincing her that $c(\mu)$ is worse than the correct action $c(\omega^*)$, there may be no signal that leads the agent to take $c(\omega^*)$. When there are two states and finite actions, always being able to displace a wrong action and always being able to take the correct action are equivalent, as they both reduce to unbounded beliefs. But more generally, it is displacing wrong actions that is fundamental for learning. 

\begin{figure}[h]
\centering
  \includegraphics[width=.6\linewidth]{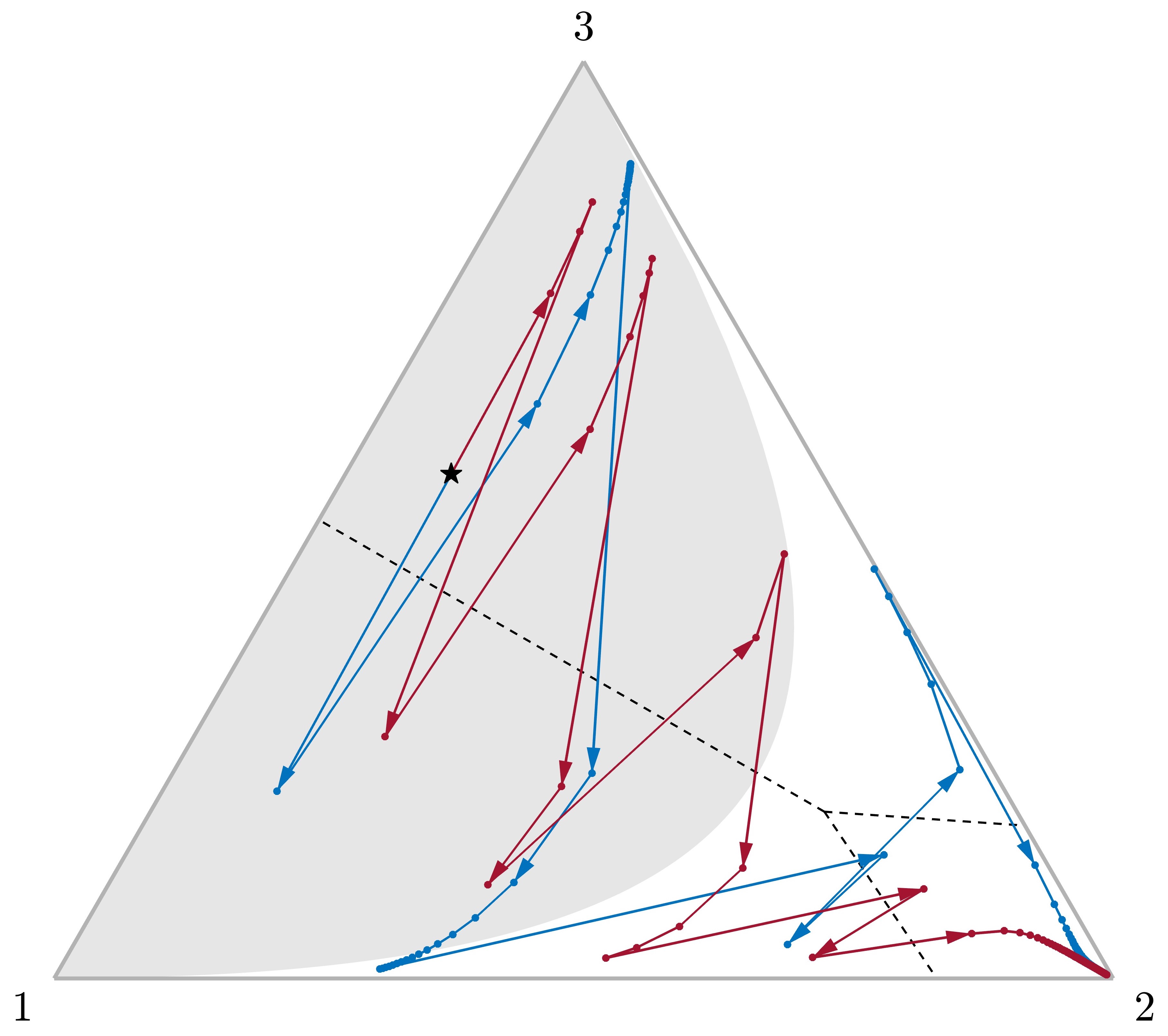}
\caption{Two simulated social belief paths---one in red and one in blue---in a complete network. There are three states labeled $1,2,3$, and there is normal information (with standard deviation $1.2$). There are three actions with respective state-contingent utilities $(1,0,-0.3)$, $(0,0.2,0)$, and $(-0.3,0,1)$. The optimal action under uncertainty is delineated by the dashed lines. The true state is $2$, and society starts with the prior $(0.35,0.1,0.55)$, marked by the black star. The grey shaded region indicates beliefs at which no single signal can lead to state 2's correct action. On each path, a dot represents the social belief after an agent has acted, and arrows indicate the sequencing.}
\label{fig:simulation}
\end{figure}

To illustrate the point concretely, consider the example depicted in \autoref{fig:simulation}. There are three states and three actions, $\Omega=A=\{1,2,3\}$. The signal structure and preferences are detailed in the figure's caption. The correct action in each state $\omega$ is $a=\omega$. Importantly, unbounded beliefs fails yet there is excludability.\footnote{Unbounded beliefs fails because under normal information the middle state is not distinguishable from its complement. Excludability can be verified by checking distinguishability of the preferred sets for each pair of actions; alternatively, we note that the preferences satisfy \hyperref[def:SCD]{single-crossing differences} (SCD), and as explained in \autoref{sec:SCD-DUB}, SCD and normal information imply excludability.}  Let agent $n$'s \emph{social belief} be her belief about the state given only the history of her neighbors' actions, prior to observing her own private signal. When each agent observes all predecessors' actions, \autoref{fig:simulation} shows two representative numerically-simulated paths of {social beliefs} given the true state $\omega=2$.  The social belief starts at the prior, marked by a star in the figure, and then evolves as agents take actions, as indicated by either of the arrowed paths. 
There is a range of beliefs, shaded in grey, such that for any social belief in that range no signal can lead an agent to take the correct action $2$. As the prior is in this range, \emph{the first agent necessarily takes a wrong action}: either $1$ (which occurs in the red path) or $3$ (the blue path). Nevertheless, even though no agent can take the correct action $2$ for a while, society never gets stuck at a wrong action: given that an agent's predecessor chose $a\in \{1,3\}$, there are signals (very high if $a=1$ and very low if $a=3$) that convince the agent that $a$ is worse than the correct action $2$, and hence the agent will not take action $a$. At some point, after enough switching between actions $1$ and $3$, the social belief is driven outside the grey region and it becomes possible for an agent to take the correct action $2$. Eventually, society settles on that action.

Turning to necessity in \autoref{thm:exclude}: for a fixed choice set, all stationary beliefs can have adequate knowledge (and hence there is adequate learning, by \autoref{lem:learning and stationary beliefs}) even absent excludability.  But when excludability fails, there is some preferred set $\Omega_{a_1,a_2}$ that cannot be distinguished from $\Omega_{a_2,a_1}$. If $\Omega$ is finite, this means that when the choice set is $
\{a_1,a_2\}$, a belief that puts small probability on $\Omega_{a_1,a_2}$ relative to $\Omega_{a_2,a_1}$ is stationary and has inadequate knowledge. 
Hence, \autoref{lem:learning and stationary beliefs} implies that excludability is necessary for learning when we seek learning for all choice sets. 
The following example illustrates these points using an infinite action set for convenience.

\begin{example}
\label{eg:responsive}
Consider $\Omega=\{0,1\}$, $A=[0,1]$, and  $u(a,\omega)=-(a-\omega)^2$. This is an example of ``responsive preferences'' \citep{Lee95,Ali18}. Fix any nontrivial signal structure and any observational network structure satisfying expanding observations.  

Adequate learning obtains by \autoref{lem:learning and stationary beliefs}, because the only stationary beliefs have certainty on one of the two states. For, given any nondegenerate belief, with positive probability the posterior-optimal action will be different from the prior-optimal action, as the uniquely optimal action equals the posterior expected state. However, excludability is equivalent to the signal structure having unbounded beliefs, as for any $a_1<a_2$, $\Omega_{a_1,a_2}=\{0\}$ and $\Omega_{a_2,a_1}=\{1\}$. So excludability is not necessary for adequate learning at choice set $A$. But absent excludability there is \emph{in}adequate learning at any non-singleton finite choice set. For, there is then some state such that any prior that puts probability close to $1$ on that state will be stationary, but this prior has inadequate knowledge.
\end{example}

The choice-set variation required by \autoref{thm:exclude} comes ``for free'' when we seek an informational condition that ensures learning for a broad-enough {class} of preferences.
Specifically, it is sufficient that for any utility function in the class and any choice set, there is another utility function that is identical on that set but makes all other actions dominated.
Since the class of all preferences has this property, and excludability for all preferences is equivalent to the information structure having unbounded beliefs, \autoref{thm:exclude} immediately implies:

\begin{corollary}
\label{cor:UBsufficient}
An information structure yields adequate learning for all preferences if and only if it has unbounded beliefs. 
\end{corollary}

This corollary extends results from the prior literature, which are either for the complete network \citep[Theorem 1]{arieli2021general} or general networks but with only two states \citep*[Theorem 2]{ADLO11}.

To our knowledge, the only prior exception to unbounded beliefs driving learning with a discrete action space is the interesting example of \citet[][Theorem 3]{arieli2021general}. They consider the complete network and a special utility function, which they call ``simple utility'', in which the payoff is $1$ if the action matches the state and $0$ otherwise. For this case, they show that \emph{pairwise distinguishability}---for any pair of states, each is distinguishable from the other---is sufficient for learning. This result also follows from \autoref{thm:exclude}; indeed, the theorem implies that learning obtains for general observational networks. For, under simple utility, 
the preferred sets for actions $a_1\neq a_2$ are just $\{a_1\}$ and $\{a_2\}$, which means 
excludability is equivalent to pairwise distinguishability.

\section{Applications}
\label{sec:applications}

Excludability permits a study of informational conditions that assure adequate learning for broad and widely-used classes of preferences. 
This section presents two such applications: one with a one-dimensional state, and one with a multi-dimensional state.


\subsection{Learning in a One-Dimensional World}
\label{sec:SCD-DUB}

In this subsection we assume a totally ordered state space: 
for simplicity, $\Omega\subset \Reals$. A function $h:\Omega \to \Reals$ is \emph{single crossing} if either: (i) for all $\omega<\omega'$, $h(\omega) > 0 \implies h(\omega') \geq 0$; or (ii) for all $\omega<\omega'$, $h(\omega) < 0 \implies h(\omega') \leq 0$. That is, a single-crossing function switches sign between strictly positive and strictly negative  at most once. 
\label{sec:conditions}

\begin{definition}
\label{def:SCD}
Preferences represented by $u:
A\times \Omega \to \Reals$ have \emph{single-crossing differences} (SCD) if for all $a$ and $a'$, the difference $u(a,\cdot)-u(a',\cdot)$ is single crossing. 
\end{definition}

SCD is an ordinal property closely related to notions in \citet{milgrom1994monotone} and \citet{athey01}, but, following \citet{KLR23}, the formulation is without an order on $ 
A$.\footnote{SCD is equivalent to there existing some order on $A$ with respect to which \citepos{athey01} ``weak single-crossing property of incremental returns'' holds.} Ignoring indifferences, SCD requires that the preference over any pair of actions can only flip once as the state changes monotonically.
SCD is widely satisfied in economic models; in particular, it is assured by supermodularity of $u$.

The key informational condition is that of distinguishing upper and lower sets from each other. More precisely, we require that for any $\omega$, $\{\omega':\omega'\ge \omega\}$ and $\{\omega':\omega'< \omega\}$ are distinguishable from each other, and $\{\omega':\omega'> \omega\}$ and $\{\omega':\omega'\le \omega\}$ are distinguishable from each other. But since a set $\Omega'$ is distinguishable from $\Omega''$ if and only if each $\omega\in \Omega'$ is distinguishable from $\Omega''$, we can simplify as follows.

\begin{definition}
\label{def:DUB}
An information structure has \emph{directionally unbounded beliefs} (DUB) 
if every $\omega$ is distinguishable from $\{\omega':\omega'<\omega\}$ and also from $\{\omega':\omega'>\omega\}$.
\end{definition}

Crucially, DUB does not require any state $\omega$ to be distinguishable from any subset of states containing both a higher and a lower state than $\omega$. 
Rather, using \autoref{lem:DUB-Distinguishability}, we can view DUB as only requiring that for any state $\omega$, there are signals that are arbitrarily more likely in $\omega$ relative to all $\omega'<\omega$, 
and also other signals that are arbitrarily more likely in $\omega$ relative to all $\omega'>\omega$. 

A leading example of DUB information is normal information. 
More generally, for any MLRP information structure, DUB can be easily checked because it reduces to pairwise distinguishability.\footnote{Regardless of the MLRP, DUB implies pairwise distinguishability. To see why the converse is true given the MLRP, consider the case of finite states. Note that for any $\omega'>\omega$, $f(s|\omega')/f(s|\omega)\to\infty$ as $s\to \sup S$ (the ratio is increasing by MLRP, and it diverges by pairwise distinguishability); similarly, the ratio goes to $0$ as $s\to \inf S$. Hence, for any $\omega'$ and $\epsilon>0$, the condition in \autoref{lem:DUB-Distinguishability} is met for $\Omega'=\{\omega'\}$ and $\Omega''=\{\omega'':\omega'' < \omega'\}$ when $S'$ is any sufficiently small upper set of signals, while for $\Omega'' = \{ \omega'':\omega'' > \omega'\}$ the condition is met when $S'$ is any sufficiently small lower set. For an infinite state space, the intuition is the same but we appeal to the monotone convergence theorem.} We note that when $A$ is finite, pairwise distinguishability is inescapable (even without the MLRP) for adequate learning in any rich-enough class of preferences.\footnote{``Rich-enough'' here means that for any two states, there is a preference in the class such that the optimal actions in those two states are disjoint.
}

Our main result in this subsection is:
\begin{proposition}\label{thm:main}
If preferences have SCD and the information structure has DUB, then there is adequate learning. Conversely, if the information structure violates DUB and there are at least two actions, then there are SCD preferences for which there is inadequate learning.
\end{proposition}

The result says that not only is DUB a sufficient informational condition for adequate learning under any SCD preferences, but it also necessary to assure learning for all SCD preferences.

Here is the logic for sufficiency. Recall that $\Omega_{a,a'}$ denotes the states in which action $a$ is strictly preferred to $a'$. SCD implies non-reversal of strict preferences: 
for any $a$ and $a'$, either $\inf \Omega_{a, a'} 
\geq \sup \Omega_{a', a}$ 
or 
$\inf \Omega_{a', a}
\geq \sup \Omega_{a,a'}$. 
DUB says 
that every upper (resp., lower) set of states and its strict lower (resp., strict upper) set are distinguishable from each other. Therefore, SCD and DUB together guarantee excludability, and so \autoref{thm:main}'s first statement follows from \autoref{thm:exclude}.

We would like to caution against the following intuition. Under SCD preferences, any inadequate-knowledge belief $\mu$ has distinct optimal actions at the extreme states of $\mu$'s support. DUB information then guarantees learning because the extreme states can be distinguished from their complements, and so $\mu$ is not stationary. While valid for finite states, this is not a generally applicable intuition. Indeed, the following example shows that pairing DUB with distinct optimal actions at all states is not a robust principle for learning.

\begin{example}
\label{eg:DUBnoSCD}
Let $\Omega=\Integers$ and $A=\Integers\cup \{a^*\}$. In any state $\omega$, the utility from any integer action $a$ is given by quadratic loss, $u(a,\omega)=-(a-\omega)^2$, whereas the action $a^*$ is a ``safe action", $u(a^*,\omega)=-\epsilon$ for a small constant $\epsilon > 0$.\footnote{Strictly speaking, quadratic-loss utility with $\Omega=\Integers$ violates our maintained assumption of bounded utility, but we ignore that to keep the example succinct.} So any action $\omega$ is uniquely optimal in state $\omega$ but worse than the safe action $a^*$ in every other state. Plainly, SCD is violated.

Consider normal information. There are full-support priors $\mu$ such that the posterior probability $\mu_s(\omega)$ is uniformly bounded away from $1$ across signals $s$ and states $\omega$ (see \hyperref[sec:eg2]{Supplementary} \autoref{sec:eg2} for details).
For any such prior, for small enough $\epsilon>0$, the safe action $a^*$ is optimal after every signal. In other words, any such prior is stationary but has inadequate knowledge. 
So \autoref{lem:learning and stationary beliefs} implies 
inadequate learning. 
\end{example}

The argument for the necessity of DUB in \autoref{thm:main} is as follows. Take any state $\omega^*$, any two actions $a_1\neq a_2$, and consider the following SCD utility: for all $\omega<\omega^*$, $u(a_1,\omega)=1$ and $u(a_2,\omega)=0$; for all $\omega\ge \omega^*$, $u(a_1,\omega)=0$ and $u(a_2,\omega)=1$; and otherwise $u(a,\omega)=-1$. 
Since all actions except $a_1$ and $a_2$ are dominated and can be ignored, \autoref{thm:exclude} implies that for there to be adequate learning, $\Omega_{a_1,a_2}=\{\omega:\omega<\omega^*\}$ and $\Omega_{a_2,a_1}=\{\omega:\omega\ge \omega^*\}$ must be distinguishable. In particular, $\omega^*$ is distinguishable from its lower set. An analogous argument shows that $\omega^*$ is distinguishable from its upper set. Since $\omega^*$ is arbitrary, DUB holds.


While our main point in this subsection is that DUB is the correct informational condition for adequate learning under SCD preferences, it is also worth noting that for any preferences violating SCD, one can show that there are DUB information structures---e.g., normal information---with inadequate learning at some choice set. In this sense SCD and DUB are a minimal pair of sufficient conditions.

\subsection{Learning in a Multi-Dimensional World} 
\label{sec:WEP-SLS}


We now turn to a multi-dimensional environment: $A,\Omega\subset \Reals^d$ for some integer $d\geq 1$.\footnote{We view any $x\in \Reals^d$ as a column vector and denote its transposition by $x'$ and its standard Euclidean norm by $\Vert x \Vert$.} 
For instance, $A=\{1,2,3\}^2$ can represent a set of feasible policies, $\Omega=\{1,2,3\}^2$ society's ideal policy, and individuals have quadratic-loss preferences $u(a,\omega)=-\Vert a-\omega \Vert^2$. Is there a natural class of information structures for which learning obtains?

More generally, consider the following class of preferences:

\begin{definition}
\label{def:HS}
Preferences 
are \emph{intermediate} if
for all $a_1\neq a_2$, either $\Omega_{a_1,a_2}=\emptyset$ or $\Omega_{a_1,a_2}=\Omega$ or there are $h\in \mathbb{R}^d$ and $c\in\mathbb{R}$ such that $\Omega_{a_1,a_2}=\{\omega: h\cdot \omega > c \}$.
\end{definition}



Introduced by \citet{Grandmont78}, intermediate preferences have preferred sets that are either trivial or half spaces; so if $\omega,\omega'\in \Omega_{a_1,a_2}$, then for any $\lambda \in (0,1)$ and $\omega''=\lambda \omega + (1-\lambda)\omega' \in \Omega$, it holds that $\omega'' \in \Omega_{a_1,a_2}$. A leading family, subsuming quadratic-loss preferences, is weighted Euclidean preferences: {$u(a,\omega)=-l((a-\omega)' W (a-\omega))$}, for some $d\times d$ symmetric positive definite matrix $W$ and strictly increasing loss function $l:\Reals_+ \to \Reals_+$.\footnote{To confirm that these are intermediate preferences, note that by simple algebraic manipulation,
$$\begin{aligned}
    & (a_1-\omega)'W(a_1-\omega) - (a_2-\omega)'W(a_2-\omega) = (a_1-a_2)'W(a_1+a_2-2\omega).
\end{aligned}$$
Hence, $\omega\in \Omega_{a_1,a_2}$ if and only if $(a_1-a_2)'W(a_1+a_2-2\omega)>0$, or equivalently, $h \cdot \omega>c$ where $h=2(a_2-a_1)'W$ and $c=(a_2-a_1)'W(a_1+a_2)$.} Another salient example, discussed by \citet[Section 5]{CN88}, is
the constant-elasticity-of-substitution utility $u(a,\omega)=\left(\sum_{i=1}^d \omega_{(i)} (a_{(i)})^r\right)^{1/r}$ where $a_{(i)}$ and $\omega_{(i)}$ denote the respective $i$-th coordinates, and $r\neq 0$ is a parameter.


Turning to information, we focus 
on the familiar class of \emph{location-shift} information structures: $S=\Reals^d$ and there is a density $g:\mathbb R^d \to \Reals_{++}$, called the \emph{standard density}, such that $f(s|\omega)=g(s-\omega)$. We restrict attention to standard densities that are uniformly continuous. The following property will be crucial.

\begin{definition}\label{def:SLS}
A location-shift information structure is \emph{subexponential} if there are $p>1$ and $M>0$ such that $g(s)< \exp(-\Vert s \Vert^p)$ for all $\Vert s \Vert>M$.
\end{definition}

A subexponential density has a thin tail in the sense that it eventually decays strictly faster than the exponential density. 
Our leading example of a subexponential location-shift information structure is multivariate normal information: there is some covariance matrix $\Sigma$ such that the distribution of signals in state $\omega$ is $\mathcal N(\omega,\Sigma)$. Here the standard density is that of $\mathcal N(0,\Sigma)$, and \autoref{def:SLS} is verified by taking any exponent $p \in (1,2)$ and any large $M>0$. Subexponential information can fail unbounded beliefs; 
for example, this is the case for normal information when $\Omega$ contains non-extreme states, i.e., there is some state in the interior of the convex hull of $\Omega$.


The main result of this subsection is:

\begin{proposition}\label{prop:HSP & SLS}
If preferences are intermediate and the information structure is subexponential location-shift, then there is adequate learning.
\end{proposition}

The result follows from \autoref{thm:exclude} and the next lemma, which says that all half spaces are distinguishable from their complements under subexponential location-shift information. Since the nontrivial preferred sets for intermediate preferences are half spaces,  the lemma implies that this combination of preferences and information yields excludability. 

\begin{lemma}\label{lem:SLS}
For a subexponential location-shift information structure, the sets $\{\omega:h\cdot \omega > c\}$ and $\{\omega:h\cdot \omega< c\}$ are distinguishable from each other for any $h\in \Reals^d$ and $c\in \Reals$.
\end{lemma}


The exponent $p$ being strictly larger than $1$ in the definition of subexponential is essential for the lemma. To see that, consider the double-exponential standard density $g(s)= c \cdot \exp(-\Vert s\Vert)$ with $c>0$ a constant of integration. This density is not subexponential, and indeed the conclusion of \autoref{lem:SLS} fails: no two states $\omega \neq \omega'$ are distinguishable from each other because $f(s|\omega')/f(s|\omega) = g(s-\omega')/g(s-\omega) \le \exp\left(\Vert \omega'-\omega\Vert\right)$ for any signal $s$. The failure of pairwise distinguishability implies inadequate learning even with a binary state when %
the action set is discrete and preferences are nontrivial.

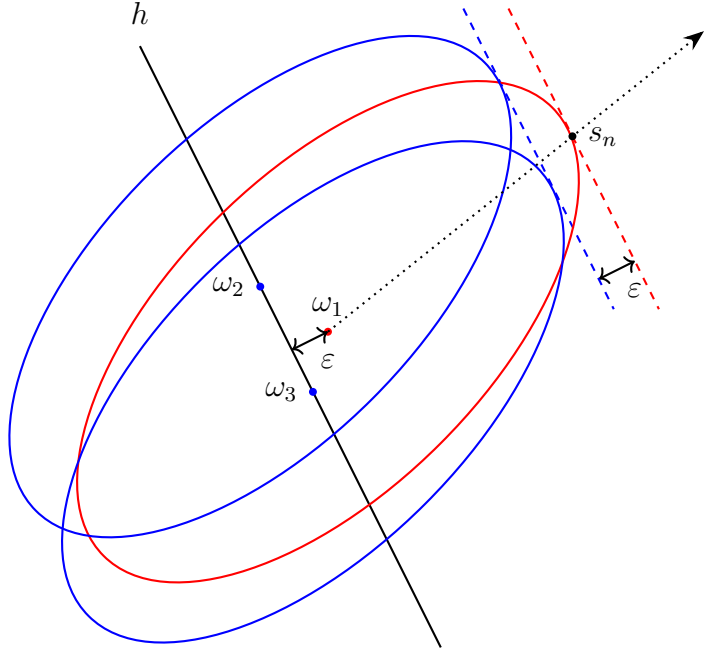
\begin{figure}[h]
    \centering 
	  \begin{tikzpicture}[scale=1]
			
		\coordinate (A) at (0,1); \coordinate (B) at (4,3); \coordinate (C) at (0,6); \coordinate (D) at (4,-2); \coordinate (Z) at (2.02,1.96);
		\coordinate (E) at (2.5,2.2); \coordinate (F) at (2.3,1.4); \coordinate (G) at (1.6,2.8);
		\coordinate (H) at (4.3,6.5); \coordinate (I) at (6.3,2.5);
		\coordinate (J) at (4.9,6.5); \coordinate (K) at (6.9,2.5); \coordinate (L) at (6.58,3.14);
		\coordinate (S) at (5.75,4.8);
		\coordinate (S2) at (6.1,2.9);
		\coordinate (S3) at (7.5,6.2);

		\draw[thick] (C) -- (D);
		\node[label={[black] above:{$h$}}] (a) at (C) {};
		\node[circle,fill=blue,inner sep=0pt,minimum size=3pt,label={[black] left:{$\omega_3$}}] (a) at (F) {};
		\node[circle,fill=red,inner sep=0pt,minimum size=3pt,label={[black] above:{$\omega_1$}}] (a) at (E) {};
		\node[circle,fill=blue,inner sep=0pt,minimum size=3pt,label={[black] left:{$\omega_2$}}] (a) at (G) {};
		
		\draw[thick,<->] (S2) -- (L);
		\node[label={[black] below:{$\e$}}] (a) at (L) {};
		\draw[thick,<->] (Z) -- (E);
		\node[label={[black] below:{$\e$}}] (a) at (E) {};
		
		\draw[blue,thick,dashed] (H) -- (I);
		\draw[red,thick,dashed] (J) -- (K);
		
		\draw[rotate=-45,red,thick] (E) ellipse (60pt and 120pt);
		\draw[rotate=-45,blue,thick] (F) ellipse (60pt and 120pt);
		\draw[rotate=-45,blue,thick] (G) ellipse (60pt and 120pt);
		
		\node[circle,fill=black,inner sep=0pt,minimum
		size=3pt,label={[black] right:{$s_n$}}] (a) at (S) {};

		\draw[black,thick,dotted,-{Stealth[scale=1.2]}] (E) -- (S3);
		
		\end{tikzpicture}
		\bigskip
	    \caption{The logic underlying \autoref{lem:SLS} for a bivariate normal standard density. We seek to distinguish $\omega_1$ from the solid black line. The ellipses are iso-density signals of a given level at the states $\omega_1$, $\omega_2$, and $\omega_3$. 
        As $s_n$ grows along the dotted line, corresponding to lower iso-density levels, $\min\{f(s_n|\omega_2)/f(s_n|\omega_1),f(s_n|\omega_3)/f(s_n|\omega_1)\} \to 0$.} 
        \label{fig:Euclidean pref}
	\end{figure}

We can provide an intuition for \autoref{lem:SLS} by considering a bivariate normal 
standard density, $g(s)=\exp\left(-s'\Sigma s/2\right)/\sqrt{2\pi}$
with $\Sigma$ a $2\times 2$ covariance matrix. Take an arbitrary hyperplane $h$, as illustrated in \autoref{fig:Euclidean pref}. We seek to distinguish the half space to the right of $h$ from its complementary half space to the left.  It is sufficient to distinguish an arbitrary single state $\omega_1$ to the right of $h$ from all the states to the left. 
\autoref{fig:Euclidean pref} shows how to construct a sequence of signals verifying that distinguishability. 
For a sequence of $c_n\to 0$, select $s_n$ on the iso-density ellipse of level $c_n$ given state $\omega_1$ so that the direction of $h$ is tangent with the ellipse at $s_n$. 
For all $n$, the ``ellipsoid distance'' between $s_n$ and $\omega_1$, $\sqrt{(s_n-\omega_1)'\Sigma(s_n-\omega_1)}$, is then smaller than the ellipsoid distance between $s_n$ and any state to the left of $h$ (such as $\omega_2$ and $\omega_3$) by some fixed amount. Due to the normal distribution being subexponential, as $c_n\to 0$ the likelihood ratio $\frac{g(s_n-\omega)}{g(s_n-\omega_1)} \to 0$ uniformly across $\omega$ to the left of $h$.

We make two further comments regarding \autoref{prop:HSP & SLS}. First, in the one-dimensional environment of \autoref{sec:SCD-DUB}, SCD is more or less equivalent to preferred sets being half spaces, 
and DUB is equivalent to the distinguishability of half spaces from their complements in the sense of \autoref{lem:SLS}. 
\autoref{prop:HSP & SLS} can thus be viewed as an extension of \autoref{thm:main} to a multi-dimensional world; the restriction to location-shift information allows us to unpack the kind of information that yields the requisite half-space distinguishability.

Second, a location-shift information structure does not have to be subexponential to guarantee learning for all intermediate preferences. But it can be shown that if the standard density $g:\Reals^d\to \Reals$ is superexponential in the sense that there are $p\in (0,1)$ and $M>0$ such that $g(s)\geq \exp(-\Vert s \Vert ^p)$ for all $\Vert s \Vert > M$, then learning fails for all nontrivial intermediate preferences when $A$ is finite. 

\section{\hyperref[lem:learning and stationary beliefs]{Theorem} \ref{lem:learning and stationary beliefs} and a General Welfare Bound}

\label{sec:cascade utility}

We now return to the general characterization of adequate learning, \autoref{lem:learning and stationary beliefs}, to explain how it is derived. The theorem is best understood as a corollary of 
a welfare bound regardless of whether there is learning.  Stating that result requires some notation. Abusing notation, let 
$$u(\mu):=\max_{a\in A}\sum_\omega u(a,\omega)\mu(\omega)$$ 
be an agent's expected utility when she takes an optimal action under belief $\mu$. Recalling that $\mu_s$ denotes the posterior given a belief $\mu$ and signal $s$, let 
$$I(\mu):= \left(\sum_{\omega\in \Omega}\int_S u(\mu_s)\dif F(s|\omega)\mu(\omega)\right)-u(\mu)$$
be the expected utility improvement from observing a private signal at belief $\mu$. Observe that $I(\mu)=0$ for any stationary belief $\mu$. We write $\Phi^{BP}\subset \Delta\Delta\Omega$ to denote the set of Bayes-plausible distributions of beliefs:  $\phi\in\Phi^{BP}$ $\iff$ $\E_\phi[\mu]=\mu_0$. Again abusing notation, we write $u(\phi):= \E_\phi[u(\mu)]$ for the expected utility of an agent under the distribution of beliefs $\phi$, and analogously write $I(\phi):=\E_\phi[I(\mu)]$. It follows that $$\Phi^S:=\left\{\phi\in \Phi^{BP}:I(\phi)=0\right\}$$ 
is the set of Bayes-plausible distributions of beliefs that are supported on the set of stationary beliefs. (We have suppressed the dependence of $\Phi^{BP}$ and $\Phi^S$ on the prior $\mu_0$.)

Building on a notion mentioned by \citet{LS15}, we can now define the \emph{cascade utility} level as
$$u_*(\mu_0):=\inf_{ \phi \in \Phi^S}u(\phi).$$
In words, $u_*(\mu_0)$ is the lowest utility level that an agent can get if her Bayes-plausible distribution of beliefs is supported on stationary beliefs. Our welfare bound 
is that eventually all agents are assured a utility level of at least $u_*(\mu_0)$. More precisely:

\begin{theorem}
\label{thm:utility diffusion}
In any equilibrium $\sigma$, $\lim \inf_n \E_{\sigma,\mu_0} [u_n] \geq u_*(\mu_0)$.
\end{theorem}



The ``if'' direction of \autoref{lem:learning and stationary beliefs} readily follows from \autoref{thm:utility diffusion}: when all stationary beliefs have adequate knowledge, a correct action is taken almost surely for any distribution of stationary beliefs, hence $u_*(\mu_0)=u^*(\mu_0)$, and we have adequate learning.

The conclusion of \autoref{thm:utility diffusion} would be straightforward if we were assured that agents eventually hold stationary beliefs. 
However, 
there are networks (with expanding observations) in which with positive probability the beliefs of an infinite number of agents are bounded away from the set of stationary beliefs;  see \autoref{exp:belief convergence} in \hyperref[sec:belief convergence]{Supplementary} \autoref{sec:belief convergence}. 

Instead, we prove \autoref{thm:utility diffusion} via an \emph{improvement principle}, as suggested by \citet{BF04} and developed by \citet*{ADLO11} and others. The foundation in our general setting is a novel compactness-continuity argument.
First, \autoref{lemma:compactness of BP} in the appendix establishes that $\Phi^{BP}$ is compact when both $\Delta \Omega$ and $\Delta \Delta \Omega$ are endowed with the Prohorov metric generated from the metric on $\Omega$.
The idea when $\Omega$ is countable is that although the prior $\mu_0$ can be supported on an infinite set, it must concentrate an arbitrarily large mass on only finitely many states. Consequently, for any $\delta>0$, there is a finite subset of states $\Omega'$ such that any Bayes-plausible distribution of beliefs puts at least $1-\delta$ probability on beliefs that put at least $1-\delta$ probability on $\Omega'$. Using Prohorov's Theorem, we then deduce that $\Phi^{BP}$ is compact.
Second, we show that the utility function $u(\phi)$ and the improvement function $I(\phi)$ are continuous (\autoref{lemma: continuous} in the appendix), and thus uniformly continuous on $\Phi^{BP}$. 


Now consider any $\varepsilon$-neighborhood of the set of Bayes-plausible distributions supported on stationary beliefs, call it $(\Phi^S)^\varepsilon$. If an agent's distribution of beliefs is in $(\Phi^S)^\varepsilon$, then her ex-ante expected utility is at least close to $u_*$, as $u(\phi)\ge u_*$ on $\Phi^S$ and $u(\phi)$ is uniformly continuous. On the other hand, if the distribution is not in $(\Phi^S)^\varepsilon$, then there is some strictly positive minimum utility improvement that the agent obtains (as the complement of $(\Phi^S)^\varepsilon$ is closed, hence compact, and $I(\phi)$ is continuous).

We can then apply an {improvement principle}. The idea is as follows, where we consider deterministic networks for simplicity.
Expanding observations guarantees that we can partition society into ``generations'' such that an agent in one generation observes a predecessor who is in either the previous generation or the current generation. We inductively argue that the lowest ex-ante utility in each generation is either close to $u_*$ or increases by a fixed amount compared to the previous generation. Consider an agent's distribution of social beliefs, $\phi$. 
Her utility $u(\phi)$ must be at least the lowest ex-ante expected utility of the previous generation, because the current agent can just mimic the agent with the largest index she observes.\footnote{With stochastic networks, the fact that an agent can obtain any observed predecessor's ex-ante expected utility through mimicking relies on our assumption that players' observation neighborhoods are drawn independently. Otherwise, whether a player has observed some predecessor may correlate with that predecessor realizing a lower utility.} Then, as explained in the previous paragraph, either $u(\phi)$ is at least close to $u_*$ (when $\phi$ is in $(\Phi^S)^\varepsilon$), or the agent can improve upon $u(\phi)$ by at least some fixed amount. Thus, the lowest ex-ante expected utility in each generation increases by a fixed amount until it becomes at least close to $u_*$. Since $\e$ was arbitrary, it follows that eventually all agents' utility must be higher than a level arbitrarily close to $u_*$, which is the conclusion of \autoref{thm:utility diffusion}.

Although previous authors have deduced versions of \autoref{lem:learning and stationary beliefs} and \autoref{thm:utility diffusion} in special environments, what allows us to establish these two general results is our novel proof methodology. We highlight two distinctions with \citet[Theorem 1]{LS15}, which is the most related existing result to \autoref{thm:utility diffusion}. 
They consider a binary-state binary-action model. In that setting, they establish a welfare bound of ``diffusion utility'',  which is the utility obtained by a hypothetical agent who observes an information structure that contains only the strongest signals (an ``expert agent'', in their terminology). 
Our cascade utility is more fundamentally tied to when learning stops, as it is defined using stationary beliefs. It is not hard to see that in general, no matter the number of states or actions, cascade utility is always at least as high as (the natural extension of) diffusion utility; \autoref{rem:diffusionutility} in \autoref{app:other} elaborates.
Typically the ranking will be strict, although \citet{LS15} note that cascade and diffusion utilities coincide in their binary-state binary-action setting.
Methodologically, \citeauthorpos{LS15} argument for a minimum improvement, like that of \citet*{ADLO11}, owes to certain monotonicity that does not extend beyond their binary-binary setting.

\begin{remark}\label{rmk:belief convergence}
Our approach to proving \autoref{thm:utility diffusion} can be adapted to 
address 
belief convergence. 
Since expanding observations is compatible with the observational network having multiple components, one cannot expect the social belief to converge even in probability.\footnote{Consider an observational network consisting of two disjoint complete subnetworks: every odd agent observes only all odd predecessors, and symmetrically for even agents. Given any specification in which learning would fail on a complete network---such as the canonical binary state/binary action herding example---there is positive probability of the limit belief among odd agents being different from that among even agents.} Furthermore, there can be a positive probability that the social belief is not eventually even in a neighborhood of the set of stationary beliefs, as already noted.
Nevertheless, there are reasonable conditions under which convergence to the stationary set does obtain.
Consider deterministic networks and assume that society can be covered by finitely many subsequences such that in each subsequence agent $n_k$ observes $n_{k-1}$. Then, denoting agent $n$'s (random) social belief by $\mu_n$, it holds that for all $\epsilon>0$, $\lim_{n\to \infty} \Pr(\mu_n\in S^\epsilon)=1$, where $S^\epsilon$ denotes the $\epsilon$-neighborhood of the set of stationary beliefs. See \autoref{prop:beliefconvergence} in \hyperref[sec:belief convergence]{Supplementary} \autoref{sec:belief convergence}. We note that this result applies, in particular, to the immediate-predecessor network and the complete network. The latter is special because the social belief is then a martingale, which is assured to converge almost surely by the martingale convergence theorem. For this case, \citet[Lemma 1]{arieli2021general} have established that the limit is stationary.
\end{remark}

\begin{remark}
\label{rem:eps-excludability}
    \autoref{thm:utility diffusion} can be used to quantify how a failure of excludability impacts welfare. \autoref{cor:eps-distinguishability} in \hyperref[sec:eps-excludability]{Supplementary} \autoref{sec:eps-excludability} provides a formal result in this vein. In particular, that result implies a sense in which an environment with ``approximate excludability'' ensures that, eventually, agents' ex-ante expected 
    utilities are close to the full-information utility.
\end{remark}

\section{Concluding Remarks}
\label{sec:conclusion}

This paper has studied a general model of sequential social learning on observational networks. Our main theme has been how learning turns jointly on preferences and information when there are multiple states. We close by commenting on certain aspects of our approach.

First, our model assumes ``non-anonymous sampling'', i.e., whenever an agent sees the action of some predecessor, she knows the identity of that predecessor. However, our methodology extends to anonymous sampling, i.e., when each agent observes only the frequencies of actions in their realized neighborhood, as in \citet{SS2020}. Our results apply in that case when expanding observations (condition \eqref{eq:exp obs}) holds for the ``induced network structure'' $(\tilde Q_n)_{n\in \Naturals}$ where each $\tilde Q_n$ is defined by first drawing a neighborhood $B_n$ according to $Q_n$ and then uniform-randomly drawing a single agent from $B_n$.  \autoref{app:backbone} (\autoref{fn:unobservable index}) explains why. Interestingly, the condition coincides with \citepos{SS2020} ``non-over-sampling'' requirement. Note that expanding observations for the induced network $(\tilde Q_n)$ is more demanding than expanding observations for $(Q_n)$; this is not surprising since agents have less information when they cannot observe identities. Nevertheless, the requirement is satisfied, for example, when each agent observes the action of a uniform-randomly drawn predecessor or the actions of all predecessors---in either case, not observing their identities. But the requirement is violated when each agent $n$ observes either agent $1$ or agent $n-1$, but doesn't observe the identity (whereas expanding observations holds here when the identity is observed).

Second, the notion of learning we have adopted considers all possible priors. While this strengthens our sufficiency results, it correspondingly weakens our necessity results. With only two states, learning at any single (nondegenerate) prior is equivalent to learning at all priors. 
Our earlier working paper \citep*[Supplementary Appendix SA.1]{KLLR22} provides some analysis concerning the extent to which this is true with multiple states.

Third, our analysis has not touched on the speed of learning/welfare convergence. For binary states and the complete network, \citet{RV19} deduce the condition on the likelihood of extreme posteriors that determines whether learning is, in certain senses, efficient; they point out that their condition is violated by normal information. See \citet{HCMT18} as well.  

Lastly, our work only addresses Bayesian learning with correctly specified agents. There is a large literature on non-Bayesian social learning, surveyed by \citet{GolubSadler16}. There has also been recent interest in (mis)learning among misspecified Bayesian agents; see, for example, 
\citet{FII20}
and \citet{BohrenHauser21}.


\bibliographystyle{ecta}
\bibliography{KLLR}

\newpage

\addtocontents{toc}{\protect\setcounter{tocdepth}{1}} 

\appendix
\noindent {\LARGE \textbf{Appendices}}

\autoref{app:backbone} contains the proofs for Theorems \ref{lem:learning and stationary beliefs}--\ref{thm:utility diffusion}. \autoref{app:applications} contains the proofs of \autoref{thm:main} and \autoref{lem:SLS} (which proves \autoref{prop:HSP & SLS}).  \autoref{app:other} contains the proof of \autoref{lem:DUB-Distinguishability} and \autoref{rem:diffusionutility} on cascade vs.~diffusion utility .

\section{Backbone Results}
\label{app:backbone}
\label{general_setup}

In this section, we prove our three theorems in the following setting, which is more general than that described in the main text:

\begin{itemize}
    \item The action space and signal space $(A,\mathcal{A}),(S,\mathcal{S})$ are standard Borel spaces;

    \item The state space $\Omega$ is equipped with a metric $d$ and its Borel sigma-algebra, $\mathcal{B}(\Omega)$, such that $(\Omega,d)$ is a sigma-compact Polish space;\footnote{That is, $(\Omega,d)$ is a complete and separable metric space that can be represented as a countable union of compact sets.}

    \item The utility function $u(a,\omega)$ has absolute value uniformly bounded by $\bar{u}$ and it is pointwise equicontinuous when regarded as a collection of functions of $\omega$ indexed by $a$; moreover, for every belief (Borel probability distribution over $\Omega$), there exists an optimal action;

    \item The information/signal structure $F(\cdot|\omega)$ is a Markov kernel from $(\Omega,\mathcal{B}(\Omega))$ to $(S,\mathcal{S})$ that is continuous in $\omega$ in the total variation (TV) sense;
    
    \item The network structure is given by $Q\equiv (Q_n)_{n\in \Naturals}$, where each $Q_n$ is a probability measure over all neighborhoods, i.e., all subsets of $\{1, 2, \dots, n-1\}$, independent across $n$, independent of the state $\omega$, and independent of any private signals.
\end{itemize}




When $\Omega$ is countable as in the main text, we endow it with the discrete metric so that the sigma-compactness and continuity requirements are trivially satisfied. 

\paragraph{Discontinuous utilities.} While we make a continuity assumption on preferences, our main results hold for utilities satisfying the following condition that permits discontinuities (cf.~ \autoref{rmk:discontinuity}): 
\begin{condition}
There is a countable partition of $\Omega$ into Borel sets $B_i$ and pointwise equicontinuous functions $v_i:\Omega \to \Reals$ uniformly bounded by $\bar u$ such that $v_i|_{B_i}=u$. 
\end{condition}
\noindent To obtain our results for such utilities, we can define a new state space $\tilde \Omega$ as a disjoint union: $\tilde \Omega:=\bigsqcup \Omega_i$, where each $\Omega_i$ is a copy of $\Omega$. Choose any metric on $\tilde \Omega$ that induces the disjoint union topology. 
Define a utility $\tilde u$ on $\tilde \Omega$ by $\tilde u|_{\Omega_i}:= v_i$ for each $i$.
It follows that $\tilde \Omega$ is sigma-compact Polish and $\tilde u$ is pointwise equicontinuous and uniformly bounded. The information structure is defined such that on each $\Omega_i$ it is the same as before. 
Using our results for the new setting, one can deduce Theorems \ref{lem:learning and stationary beliefs}--\ref{thm:utility diffusion} for the original setting.\footnote{More specifically, the results in the original setting are equivalent to the corresponding results in the new setting restricted to priors/beliefs that put zero probability on the added states $\Omega_i\backslash B_i$. We can use our methodology to derive Theorems \ref{lem:learning and stationary beliefs}--\ref{thm:utility diffusion} in the new setting for such restricted beliefs.}

\subsection{Overarching Probability Space and Beliefs}

We now formalize the overarching probability space over all realizations of the state, signals, observation neighborhoods, and actions. We also define formal objects corresponding to 
agents' social and posterior beliefs and distributions of beliefs. 

\paragraph{Overarching probability space.} Our probability space is constructed from three components: the Markov kernel $F$ and probability space $(\Omega,\mathcal{B}(\Omega),\mu_0)$; the network structure $Q\equiv(Q_n)_{n \in \Naturals}$; each agent $n$'s strategy $\sigma_n(\cdot|a_{B_n},B_n)$ as a Markov kernel from $(A^{|B_n|},\mathcal{A}^{|B_n|})$ to $(A,\mathcal{A})$ for each realization of neighborhood $B_n$.

Taken together, for the first $n$ agents, we can define a probability space that describes the joint distribution of their neighborhoods, signals, actions, and the states. Since all these elements lie in standard Borel spaces, the Kolmogorov Extension Theorem guarantees existence of an overarching probability space $(H_\infty,\mathcal{H}_\infty,\P)$ that is consistent with each finite probability space (i.e., up to each agent $n$). We suppress the dependence of $\P$ on $\sigma$ and $\mu_0$.

\paragraph{Beliefs.} Given this overarching probability space, agent $n$'s social belief (i.e., her belief after observing her neighbors and their actions, but before observing her private signal) is $\P(\cdot|a_{B_n},B_n)$ and her posterior belief is $\P(\cdot|a_{B_n},B_n,s_n)$. These beliefs are well-defined because, as a countable product of standard Borel spaces, the overarching probability space is a standard Borel space, and hence there exist regular conditional probabilities \citep[Theorem 4.1.17]{durrett2019probability}. 

\paragraph{Distribution of beliefs.} We denote by $\Delta \Omega$ the space of beliefs (Borel probability measures on $\Omega$) equipped with the Prohorov metric, and by $\Delta\Delta \Omega$ the space of belief distributions (Borel probability measures on $\Delta \Omega$) also equipped with the Prohorov metric.

The social belief of agent $n$, $\mu_n$, as a regular conditional probability, can be regarded as a measurable function from $(H_\infty,\mathcal{H}_\infty,\P)$ to $(\Delta\Omega,\mathcal{B}(\Delta\Omega))$; see \citet[Remark 3.20]{crauel2002random}. As $\Omega$ is a Polish space, so is $\Delta\Omega$. We define agent $n$'s distribution of social beliefs, $\phi_n$, as the push-forward measure of $\mu_n$. Hence, $\phi_n\in \Delta\Delta \Omega$ since it is by definition a Borel probability measure on $\Delta\Omega$.


\subsection{Space of Bayes-Plausible Belief Distributions is Compact}

Given a prior $\mu_0\in \Delta\Omega$ and a
strategy profile $\sigma$, any agent's belief distribution $\phi \in \Delta \Delta \Omega$ must be Bayes plausible:
\begin{equation}\label{eq: bayes-consistency}
    \int_A \mu(A)\dif \phi(\mu)=\mu_0(A),\ \forall A\in \mathcal{B}(\Omega).
\end{equation}
Let $\Phi^{BP}\subset \Delta\Delta \Omega$ be the set of Bayes-plausible belief distributions; note that we suppress the dependence of $\Phi^{BP}$ on $\mu_0$.

Our goal is to establish (\autoref{lemma:compactness of BP} below) that even though the set of belief distributions $\Delta \Delta \Omega$ need not be compact, the subset of Bayes-plausible distributions $\Phi^{BP}$ is. A key step is the following lemma, which shows that any belief distribution $\phi\in \Phi^{BP}$ has to put a large probability on a compact subset of $\Delta\Omega$. 

\begin{lemma}\label{lemma: compact of V}
Let $\delta>0$ and $\{\Omega_i\}_{i\in \Naturals}$ be a sequence of compact sets with $\mu_0(\Omega_i)\ge 1-(\frac{\delta}{2^i})^2$,  $\forall i$. Defining
$V_\delta:=\{\mu\in \Delta\Omega:\mu(\Omega_i)\ge 1-\frac{\delta}{2^i},\ \forall i\}$, it holds that:
\begin{enumerate}
    \item $V_\delta$ is compact;

    \item $\phi(\mu\notin V_\delta)<\delta$, $\forall \phi \in \Phi^{BP}$.
\end{enumerate}
\end{lemma}
Intuitively, in the lemma's statement, the set $V_{\delta}$ contains all beliefs that put high probability on a set of states that the prior $\mu_0$ ascribes high probability to. The lemma concludes that the set $V_\delta$ is compact and that any Bayes-plausible belief distribution must put high probability on $V_{\delta}$.

\begin{proof}
(\underline{Part 1}) First, $V_\delta$ is closed. To see this, take any $\mu_k\rightarrow \mu$ and $\mu_k\in V_\delta$. Since each $\Omega_i$ is compact (and thus closed), weak convergence implies
$$\limsup_k \mu_k(\Omega_i)\le \mu(\Omega_i),\ \forall i,$$
which implies $\mu(\Omega_i)\ge 1-\frac{\delta}{2^i}$. Thus, $\mu\in V_\delta$, and hence $V_\delta$ is closed.

Next, the beliefs in $V_\delta$ are tight by definition. Hence, by Prohorov's theorem, the closure of $V_\delta$, which is $V_\delta$ itself, is compact.

(\underline{Part 2}) Note that $\phi(\mu \notin V_\delta)=\phi(\cup_i\{\mu(\Omega_i^c)> \frac{\delta}{2^i}\})\le \sum_i\phi(\mu(\Omega_i^c)> \frac{\delta}{2^i})$. For each $i \in \Naturals$, we view $\mu(\Omega_i^c)$ as a non-negative random variable 
with distribution induced by $\phi$. Since $\phi$ is Bayes plausible, $\E_\phi[\mu(\Omega_i^c)]=\mu_0(\Omega_i^c)\le (\frac{\delta}{2^i})^2$, which implies (using Markov's inequality) that $\phi(\mu(\Omega_i^c)> \frac{\delta}{2^i})<\frac{\delta}{2^i}$. This implies that $\phi(\mu\notin V_\delta) < \sum_i \frac{\delta}{2^i} = \delta$.
\end{proof}

Given \autoref{lemma: compact of V}, 
we can use Prohorov's theorem again to show:
\begin{lemma}\label{lemma:compactness of BP}
$\Phi^{BP}$ is compact.
\end{lemma}
\begin{proof}
First, we prove that $\Phi^{BP}$ is closed. Take any $\phi_k\rightarrow \phi$ and $\phi_k\in \Phi^{BP}$, and want to show that $\phi\in \Phi^{BP}$, i.e., $\E_\phi[\mu(W)]=\mu_0(W),\forall W\in \mathcal{B}(\Omega)$.

Take any open set $W \in \mathcal{B}(\Omega)$. For any $\mu_k\rightarrow \mu$, it holds that $\mu(W) \leq \liminf \mu_k(W)$. In other words, $\mu(W)$ (as a function of $\mu$) is lower semi-continuous. By properties of weak convergence, it follows that $\E_{\phi}[\mu(W)]\le\liminf \E_{\phi_k}[\mu(W)]=\mu_0(W) $. That is, the mean measure of $\phi$ ascribes a smaller probability than $\mu_0$ to any open set.

Now observe that $W^c \subseteq \cup_{x\in W^c}B_{1/n}(x)$ for any $n$. Hence, 
$$ 
\E_{\phi}[\mu(W^c)] \leq \lim_n \E_{\phi}[\mu(\cup_{x\in W^c}B_{1/n}(x))] \leq \lim_n \mu_0(\cup_{x \in W^c} B_{1/n}(x)) = \mu_0(W^c),
$$
where the second inequality is from the previous result applied to open sets $\cup_{x\in W^c}B_{1/n}(x)$, and the last equality follows from $W^c=\cap_n\cup_{x\in W^c}B_{1/n}(x)$ (and this equality holds because $W^c$ is closed). Therefore, $\E_{\phi}[\mu(W)] = \mu_0(W)$.

Since $\E_{\phi}[\mu]$ and $\mu_0$ agree on all open sets, and open sets generate $\mathcal{B}(\Omega)$, $\E_\phi[\mu]$ and $\mu_0$ agree on all sets in $\mathcal{B}(\Omega)$. This establishes that $\phi \in \Phi^{BP}$.

Finally, $\Omega$ being sigma-compact implies that for any $\delta$, there is an increasing sequence of compact sets $\{\Omega_i\}_{i\in\Naturals}$ such that $\Omega=\cup_{i} \Omega_i$, and this sequence $\{\Omega_i\}$ satisfies the hypotheses in \autoref{lemma: compact of V}. The lemma guarantees that there is a compact set $V_\delta$ such that $\phi(V_\delta)<\delta$ for all $\phi\in \Phi^{BP}$, hence $\Phi^{BP}$ is tight. Prohorov's theorem now implies that the closure of $\Phi^{BP}$, which is $\Phi^{BP}$ itself, is compact.
\end{proof}

\subsection{Continuity of Various Functions}
We next define some functions of interest, some of which were already defined in the main text but are now defined for the more general setting considered in the appendix. 

Let  $u(\mu)$ be the expected utility that an agent can get at belief $\mu$:
$$u(\mu):=\sup_{a\in A} \int_\Omega u(a,\omega)\dif \mu(\omega).$$
Let $u^F(\mu)$ be the expected utility that an agent can get at belief $\mu$, if she can choose an action after observing her private signal:
$$u^F(\mu):=\sup_{\beta: S \to A}\int_\Omega\int_S u(\beta(s),\omega)\dif F(s|\omega)\dif \mu(\omega).$$
Finally, let $u^*(\mu)$ be the full information utility at $\mu$:
$$u^*(\mu):=\int_\Omega \sup_{a\in A} u(a,\omega)\dif \mu(\omega).$$

Our continuity assumptions on the utility function and the information structure allow us to prove:
\begin{lemma}\label{lemma: continuous}
$u,u^F,u^*$ are continuous in $\mu$.
\end{lemma}

To prove \autoref{lemma: continuous}, we use Theorem 2.2.8 in \cite{bogachev2018weak}, which we restate without proof for our context as the following claim:
\begin{claim}\label{lemma: Theorem 2.2.8 in B2018}
Let $\mu_k \to \mu$. If $\Gamma$ is a uniformly bounded and pointwise equicontinuous family of functions on $\Omega$, then
$$\lim_k \sup_{f\in \Gamma}\left|\int_\Omega f\dif \mu_k-\int_\Omega f\dif \mu\right|=0.$$
\end{claim}

\begin{proof}[Proof of \autoref{lemma: continuous}]
By assumption, $\Gamma := \{u(a, \omega)\}_{a \in A}$, viewed as a family of functions of $\omega$ indexed by $a$, is uniformly bounded and pointwise equicontinuous.

Consider the function $u^*$. Since the supremum of the pointwise equicontinuous functions $u^*(\omega) := \sup_{a} u(a, \omega)$ is continuous in $\omega$, the definition of weak convergence implies that $u^*(\mu)$ is continuous in $\mu$.

Now consider the function $u$. Its continuity follows from
$$|u(\mu_k) - u(\mu)| = \left|\sup_{f\in \Gamma}\int_\Omega f\dif \mu_k -\sup_{f\in \Gamma}\int_\Omega f\dif \mu \right|\le \sup_{f\in \Gamma}\left|\int_\Omega f\dif \mu_k-\int_\Omega f\dif \mu\right|,$$
which converges to $0$ as $\mu_k \to \mu$, by  \autoref{lemma: Theorem 2.2.8 in B2018}.

Lastly, suppose we establish that  $\Gamma^F:= \left\{\int_S u(\beta(s),\omega)\dif F(s|\omega)\right\}_{\beta: S \to A}$, as a family of functions of $\omega$ indexed by $\beta$, is 
pointwise equicontinuous.\footnote{Here we assume $\beta$ are (measurable) pure strategies for notation clarity. The same argument works for mixed strategies, in which case $\beta$ would be Markov kernels.} Then, as $\Gamma^F$ is  uniformly bounded, \autoref{lemma: Theorem 2.2.8 in B2018} implies that $u^F(\mu)$ is continuous, proving the lemma.

To establish the pointwise equicontinuity of $\Gamma^F$, observe that $\forall \omega, \omega'$ and $\forall \beta,$ 
\begin{align}\label{eqn:equicontinuity_gammaF}
& \left|\int_S u(\beta(s),\omega)\dif F(s|\omega) - \int_S  u(\beta(s),\omega^\prime)\dif F(s|\omega^\prime)\right|\\
\le & \left|\int_S (u(\beta(s),\omega)-u(\beta(s),\omega^\prime))\dif F(s|\omega) \right| \notag
+\left|\int_S u(\beta(s),\omega^\prime)\dif F(s|\omega) - \int_S u(\beta(s),\omega^\prime)\dif F(s|\omega^\prime)\right|.
\end{align}
Fix any $\omega$ and any $\epsilon > 0$. Since $\{u(a, \omega)\}_{a \in A}$ is pointwise equicontinuous, there exists $\delta_1$ such that $d(\omega', \omega) < \delta_1$ implies the first term on the right-hand side of inequality \eqref{eqn:equicontinuity_gammaF} to be smaller than $\epsilon/2$ (regardless of $\beta(s)$). The second term is smaller than $2\bar{u}d_{TV}(F(\cdot|\omega),F(\cdot|\omega^\prime))$ (where TV represents total variation), and by the continuity assumption of the information structure, there exists $\delta_2 > 0$ such that $d(\omega', \omega) < \delta_2$ implies $d_{TV}(F(\cdot|\omega),F(\cdot|\omega^\prime)) < \epsilon/4\bar{u}$. Therefore, if $d(\omega', \omega) < \min\{\delta_1, \delta_2\}$, then the right-hand side of inequality \eqref{eqn:equicontinuity_gammaF} is less than $\epsilon$ (regardless of $\beta(s)$). It follows that $\Gamma^F$ is pointwise equicontinuous. 
\end{proof}

Now define the utility improvement $I(\mu)$ and the utility gap $G(\mu)$ at $\mu$ as:
$$I(\mu):=u^F(\mu)-u(\mu),\quad G(\mu) := u^*(\mu)-u(\mu).$$
By \autoref{lemma: continuous}, $I(\mu)$ and $G(\mu)$ are continuous. Lastly, with an abuse of notation, define $u(\phi):=\E_\phi[u(\mu)]$,  $I(\phi):=\E_\phi[I(\mu)]$, and $G(\phi):=\E_\phi[G(\mu)]$ as the corresponding functions over distributions of beliefs. Since $u(\mu)$, $I(\mu)$, and $G(\mu)$ are continuous, so are $u(\phi)$, $I(\phi)$, and $G(\phi)$.

We note that a belief $\mu$ is stationary if and only if $I(\mu)=0$, and a belief $\mu$ has adequate knowledge if and only if $G(\mu)=0$. To confirm these points, consider stationary beliefs. If there is an action that is a.s.~optimal regardless of the signal, then  $I(\mu)=0$. Conversely, if no action is a.s.~optimal regardless of the signal, then for any action there is a positive-probability set of signals for which that action is strictly suboptimal; hence $u^F(\mu)>u(\mu)$, and $I(\mu)>0$. The argument for adequate knowledge beliefs is similar.

\subsection{Proofs for Backbone Results}

Logically, \autoref{thm:utility diffusion} $\implies$ \autoref{lem:learning and stationary beliefs} $\implies$ \autoref{thm:exclude}. So we prove the results in that order.

\begin{proof}[Proof of \autoref{thm:utility diffusion}]

We prove the result in two steps. In Step 1 below, we prove that if agent $n$'s social belief distribution $\phi_n$, which is her belief distribution incorporating the observation of her neighborhood's actions but not her private signal, is not close to being supported on only stationary beliefs, then her utility $\E_{\sigma, \mu_0} [u_n]$, which is the ex-ante expected utility under equilibrium $\sigma$ after observing the private signal, improves from $u(\phi_n)$ by some positive amount bounded away from zero. In Step 2 below, we use the expanding observations assumption to establish that this minimum improvement propagates through the network until eventually agents obtain at least arbitrarily close to their cascade utility level.

\underline{Step 1}: Recall the set of Bayes-plausible belief distributions that are supported by stationary beliefs, $\Phi^S:=\{\phi\in \Phi^{BP}:I(\phi)=0\}$, and the \emph{cascade utility}, $u_*:=\inf_{\phi\in \Phi^S}u(\phi)$.

Take any $\varepsilon>0$, and let $(\Phi^S)^\e$ denote the $\varepsilon$-neighborhood of $\Phi^S$. 
An agent $n$'s belief distribution $\phi_n$ must be Bayes plausible, so $\phi_n \in \Phi^{BP}$.
Since $u(\phi)$ is uniformly continuous on $\Phi^{BP}$ (as $u(\phi)$ is continuous, and $\Phi^{BP}$ is compact), if $\phi_n \in (\Phi^S)^\e$, 
then $u(\phi_n)\ge u_*-\gamma(\varepsilon)$ for some $\gamma(\cdot)$ such that $\gamma(\varepsilon)\rightarrow 0$ when $\varepsilon\rightarrow 0$. If, on the other hand, $\phi_n \in \Phi^+ := \Phi^{BP} \backslash (\Phi^S)^{\varepsilon}$, then $I(\phi_n)>0$ because $\phi_n$ puts positive probability on $\{\mu:I(\mu)>0\}$. Since $(\Phi^S)^{\varepsilon}$ is open, $\Phi^{+}$ is a closed subset of a compact set $\Phi^{BP}$; hence $\Phi^{+}$ is compact, and since $I(\phi)$ is continuous, it attains a minimum over $\Phi^+$ at some $\underline{\phi}\in \Phi^+$. Thus, if $\phi_n \in \Phi^+$ the agent obtains an improvement $I(\phi_n) \geq I(\underline{\phi})>0$.

\underline{Step 2:} We will argue that for any $\epsilon>0$, $\E_{\sigma, \mu_0} [u_n] \ge u_*-\gamma(\varepsilon)$ once $n$ is large enough. Since $\varepsilon$ is arbitrary, taking $\varepsilon \to 0$ implies $\liminf_n \E_{\sigma, \mu_0} [u_n] \ge u_*$, which completes the proof.

For a given $\epsilon>0$, let $\delta= \frac{I(\underline{\phi})}{4\bar{u}} > 0$, let $N_0=1$, and define $N_k$ for $k = 1, 2, \dots$ sequentially such that for all $n\ge N_k$, $Q_n(\max_{b\in B_n}b<N_{k-1})<\delta$. Expanding observations ensures that such $N_k$ exist. 

We claim that, for any agent $n \geq N_k$, $\E_{\sigma, \mu_0} [u_n] \geq \alpha_k := \min\{u_* - \gamma(\varepsilon), \frac{k I(\underline{\phi})}{2} - \bar{u}\}$. 
Since $\alpha_0=-\bar u$, clearly $\E_{\sigma, \mu_0} [u_n] \geq \alpha_0$ for any $n \geq N_0$. Suppose the claim holds for all agents $n' \geq N_{k-1}$. Take any agent $n \geq N_{k}$. Agent $n$'s neighborhood is drawn independently of everything that has happened before, so conditional on agent $n$ observing an agent $n' \geq N_{k-1}$, even without her private signal agent $n$ can achieve a utility of at least $\alpha_{k-1}$ by imitating agent $n'$. Hence, $u(\phi_n) \geq (1-\delta) \cdot \alpha_{k-1} + \delta \cdot (-\overline{u})$.\footnote{\label{fn:unobservable index} 
If agents do not observe the identities associated with the observed actions of their predecessors, an agent can uniform-randomly select one of the actions they observe to imitate. So long as the ``induced network structure'' (i.e., a network structure $(\tilde Q_n)$ wherein each $\tilde Q_n$ is defined by first drawing a neighborhood $B_n$ from $Q_n$ and then uniform-randomly drawing a single agent from $B_n$) satisfies expanding observations, the current proof goes through without change using the induced network structure.} 
If $\phi_n\in (\Phi^S)^\e$, then by definition $u(\phi_n) \ge u_*-\gamma(\epsilon)$, and thus $\E_{\sigma, \mu_0} [u_n] \geq u(\phi_n) \ge u_* - \gamma(\varepsilon) \geq \alpha_{k}$. If $\phi_n\notin (\Phi^S)^\e$, then agent $n$ can improve her utility by at least $I(\underline{\phi})$, and so
$$
\begin{aligned}
\E_{\sigma, \mu_0} [u_n] & \geq (1-\delta) \alpha_{k-1} + \delta \cdot(-\overline{u}) + I(\underline{\phi})\\
& \ge  \alpha_{k-1} + \frac{I(\underline{\phi})}{2} \quad \text{(because $\alpha_{k-1} \leq \overline{u}$ and $\delta = \frac{I(\underline{\phi})}{4\bar{u}}$)}\\
& \geq \alpha_k.
\end{aligned}
$$

Since the definition of $\alpha_k$ implies that there is a finite $K$ such that for all $k\geq K$, $\alpha_k= u_*-\gamma(\e)$, it follows that for all $n\geq N_K$, $\E_{\sigma, \mu_0} [u_n] \ge u_*-\gamma(\varepsilon)$.
\end{proof}

\begin{proof}[Proof of \autoref{lem:learning and stationary beliefs}]
The ``only if'' direction is straightforward. If there is a stationary belief without adequate knowledge, then when the prior is that belief there is an equilibrium where each agent ignores her signal and action history and obtains a utility that is strictly below the full-information utility level.

For the "if" direction, fix any prior $\mu_0$ and equilibrium $\sigma$. Since all stationary beliefs have adequate knowledge, $I(\mu)=0$ implies $G(\mu)=0$. Thus, for any $\phi\in \Phi^S$, $\phi(\{\mu:I(\mu)=0\})=\phi(\{\mu:G(\mu)=0\})=1$, which implies $G(\phi)=u^*(\phi)-u(\phi)=0$. Moreover, because $\mu_0$ is the mean measure of $\phi$,
$$u^*(\phi)=\E_\phi\left[\int_\Omega \sup_a u(a,\omega)\dif \mu\right]=\int_\Omega \sup_a u(a,\omega)\dif \mu_0=u^*(\mu_0),$$
which implies $u(\phi)=u^*(\mu_0)$. As a result, $u_*(\mu_0)=\inf_{\phi \in \Phi^S} u(\phi)=u^*(\mu_0)$. It follows from \autoref{thm:utility diffusion} that $\liminf_n\E_{\sigma,\mu_0}[u_n]\ge u^*(\mu_0)$. Since $\E_{\sigma,\mu_0}[u_n]\le u^*(\mu_0)$ for all $n$, it further follows that $\E_{\sigma,\mu_0}[u_n]\rightarrow u^*(\mu_0)$. As $\mu_0$ and $\sigma$ are arbitrarily, we have adequate learning.
\end{proof}

Next we state and prove a more general version of \autoref{thm:exclude}. For any $n\in \Naturals$, define $\Omega_{a_1,a_2}^n:=\{\omega:u(a_1,\omega)-u(a_2,\omega)>\frac{1}{n}\}$.

\begin{thmbis}{thm:exclude}\label{exclude_prime}
Excludability implies adequate learning at every choice set.
There is inadequate learning for choice set $\{a_1,a_2\}$ if $\Omega_{a_1,a_2}$ is not distinguishable from $\Omega_{a_2,a_1}^n$ for some $n$.
\end{thmbis}

Note that when $\Omega$ is finite, or the utility difference between any pair of actions is bounded away from zero, a failure of excludability is equivalent to the condition for necessity in the theorem holding for some $a_1,a_2$. Hence \autoref{thm:exclude} is implied by \autoref{exclude_prime}.

\begin{proof}[Proof of \autoref{exclude_prime}]
(\underline{First statement}) First note that excludability (under the full choice set $A$) implies excludability under any choice subset $A'\subseteq A$. 
So we fix an arbitrary $A'\subseteq A$ and show that excludability under that subset implies adequate learning at that choice subset. In what follows, the domain of actions should be understood as $A'$, and we denote a typical element by $a'$.

\autoref{lem:learning and stationary beliefs} implies that we need only show that any $\mu\in \Delta\Omega$ with inadequate knowledge is not stationary. So take any $\mu \in \Delta \Omega$ with inadequate knowledge and any $a^* \in c(\mu)$. Since there is inadequate knowledge, $\mu(\cup_{a'}\Omega_{a',a^*})>0$, i.e., there is a positive measure of states where $a^*$ is not optimal. The continuity of $u(a',\omega)-u(a^*, \omega)$ implies that $\Omega^n_{a',a^*}$ are open sets for any $a'$ and $n$. Since $\Omega$ is Polish, it is second-countable and hence has a countable basis. Therefore, each open set $\Omega^n_{a',a^*}$, and hence the open set $\cup_{a'}\Omega_{a',a^*}(= \cup_{a'} \cup_n \Omega^n_{a',a^*})$, is a union of countably many basic open sets. Since $\mu(\cup_{a'}\Omega_{a',a^*})>0$, at least one basic open set contained in $\Omega^n_{a',a^*}$ for some $a'$ and $n$ has strictly positive measure, i.e., $\mu(\Omega_{a',a^*}^n)>0$.

Now denote $\mu'(\cdot):=\mu(\cdot|\Omega_{a^*,a'}\cup \Omega_{a',a^*}^n)$ as the corresponding conditional probability. 
Since $\Omega_{a',a^*}$ is distinguishable from $\Omega_{a^*,a'}$ by excludability, so is $\Omega_{a',a^*}^n$.\footnote{In fact, excludability is equivalent to: 
$\Omega_{a_1,a_2}^n$ is distinguishable from  $\Omega_{a_2,a_1}$ for all $a_1,a_2$ and $n$.} Therefore, for any $\e>0$ there exists a set of signals $S'$ such that $\Pr_{\mu'}(S')>0$ and $\mu'_s(\Omega_{a',a^*}^n)>1-\e$ for all $s\in S'$. The utility improvement upon observing any $s \in S'$ by switching from $a^*$ to $a'$ is therefore bounded below by $(\frac{1}{n}(1-\e)-2\bar{u}\e)\mu_s(\Omega_{a^*,a'}\cup \Omega_{a',a^*}^n)$, as the expected improvement on $\Omega\backslash(\Omega_{a^*,a'}\cup \Omega_{a',a^*}^n)$ is nonnegative. For small $\e>0$, $\frac{1}{n}(1-\e)-2\bar{u}\e>0$. Furthermore, integrating $\mu_s(\Omega_{a^*,a'}\cup \Omega_{a',a^*}^n)$ over $s\in S'$ yields $\Pr_{\mu'}(S') \mu(\Omega_{a^*,a'}\cup \Omega_{a',a^*}^n)>0$. Hence, the ex-ante improvement is bounded below by $(\frac{1}{n}(1-\e)-2\bar{u}\e)\Pr_{\mu'}(S') \mu(\Omega_{a^*,a'}\cup\Omega_{a',a^*}^n)>0$. It follows that $I(\mu)>0$, and thus $\mu$ is not stationary.

(\underline{Second statement}) Suppose there are two actions $a_1,a_2$ and an $n$ such that $\Omega_{a_1,a_2}$ is not distinguishable from $\Omega_{a_2,a_1}^n$. This means there exists $\mu\in \Delta(\Omega_{a_1,a_2}\cup \Omega_{a_2,a_1}^n)$ with $\mu(\Omega_{a_1,a_2})>0$ such that $\mu_s(\Omega_{a_1,a_2})\le 1-\e$ for some $\e>0$ and $\mu$-a.e. $s$. Consider $\mu'\in \Delta(\Omega_{a_1,a_2}\cup \Omega_{a_2,a_1}^n)$ with a small 
$\mu'(\Omega_{a_1,a_2})>0$ such that $\mu'(\cdot|\Omega_{a_1,a_2})=\mu(\cdot|\Omega_{a_1,a_2})$ and $\mu'(\cdot|\Omega_{a_2,a_1}^n)=\mu(\cdot|\Omega_{a_2,a_1}^n)$. Under $\mu'$, upon observing signal $s$, the posterior on $\Omega_{a_1,a_2}$ satisfies
$$\frac{\mu'_s(\Omega_{a_1,a_2})}{\mu'_s(\Omega_{a_2,a_1}^n)}=\frac{\mu_s(\Omega_{a_1,a_2}) / \mu(\Omega_{a_1,a_2})}{\mu_s(\Omega_{a_2,a_1}^n) / \mu(\Omega_{a_2,a_1}^n)}\frac{\mu'(\Omega_{a_1,a_2})}{\mu'(\Omega_{a_2,a_1}^n)} \le \frac{1-\e}{\e}\frac{\mu(\Omega_{a_2,a_1}^n)}{\mu(\Omega_{a_1,a_2})}\frac{\mu'(\Omega_{a_1,a_2})}{\mu'(\Omega_{a_2,a_1}^n)}$$
for $\mu$-a.e.~$s$. Hence, by choosing $\mu'$ so that $\frac{\mu'(\Omega_{a_1,a_2})}{\mu'(\Omega_{a_2,a_1}^n)}$ is arbitrarily small, the ratio $\frac{\mu'_s(\Omega_{a_1,a_2})}{\mu'_s(\Omega_{a_2,a_1}^n)}$  can be made arbitrarily small uniformly over $s$.

Under $\mu'$, after observing $s$, the expected improvement by switching from $a_2$ to $a_1$ is bounded above  by $2\bar{u}\mu'_s(\Omega_{a_1,a_2})-\frac{1}{n}\mu'_s(\Omega_{a_2,a_1}^n)$, which is strictly negative when $\frac{\mu'_s(\Omega_{a_1,a_2})}{\mu'_s(\Omega_{a_2,a_1}^n)}$ is small. Therefore, for $\mu'$-a.e.~$s$, $a_2$ is strictly better than $a_1$, and thus $\mu'$ is stationary for choice set $\{a_1,a_2\}$. However, since $\mu'(\Omega_{a_1,a_2})>0$, the belief $\mu'$ has inadequate knowledge. \autoref{lem:learning and stationary beliefs} implies there is inadequate learning for choice set $\{a_1,a_2\}$.
\end{proof}

\section{Applications}\label{app:applications}

We now specialize to the main text's setting: 
$\Omega$ is countable, endowed with the discrete metric, and $F(\cdot|\omega)$ are absolutely continuous with respect to each to other, and so there are densities $f(\cdot|\omega)>0$.

\subsection{SCD Preferences \& DUB Information}
\begin{proof}[Proof of \autoref{thm:main}]

Sufficiency follows directly from \autoref{thm:exclude}. For necessity, first observe that if the information structure fails DUB, then there exists some state $\omega^*$ such that $\omega^*$ is not distinguishable from its lower set (or from its upper set, which has a symmetric argument). Fix any pair of distinct actions $a_1$ and $a_2$, and define the following SCD preferences: for $\omega<\omega^*$, $u(a_1,\omega)=1$ and $u(a_2,\omega)=0$; for $\omega\geq \omega^*$, $u(a_1,\omega)=0$ and $u(a_2,\omega)=1$; and any other actions are strictly dominated. It follows that $\Omega_{a_2,a_1}$ is not distinguishable from $\{\omega:u(a_1,\omega)-u(a_2,\omega)>\frac{1}{2}\}$. By \autoref{exclude_prime}, there is inadequate learning when the choice is $\{a_1,a_2\}$, and since all other actions are strictly dominated, also for the full choice set $A$.
\end{proof}

\subsection{Intermediate  Preferences \& Location-Shift Information}\label{sec:omitted proofs}

The proof of \autoref{lem:SLS} is more involved than the intuition given in the main text using \autoref{fig:Euclidean pref}, because in general one cannot explicitly identify the sequence of signals that establishes distinguishability of the relevant two sets.

We will use the following claim in proving \autoref{lem:SLS}. For any $h, x \in \Reals^d$, let $\Vert x \Vert_h:=h\cdot x $ be the ``signed distance'' of $x$ in direction $h$, i.e., between $x$ and the hyperplane $\{z : h \cdot z = 0\}$. 
Note that $\Vert\cdot\Vert_h$ is linear, so $\Vert x-x'\Vert_h=\Vert x\Vert_h-\Vert x'\Vert_h$.

\begin{claim}
\label{claim:strictly subexponential}
If a standard density $g$ is subexponential, then for any $\overline s$ with $\Vert\bar{s}\Vert_h>0$, and $\epsilon \in (0,1)$, there is $s$ with $\Vert s-\bar{s}\Vert_h\ge 1$ such that:
\begin{enumerate}
    \item $\sup_{\{s':\Vert s'-s\Vert_h\ge 1/\Vert\overline s\Vert_h\}}\frac{g(s')}{g(s)} < \epsilon$; and 
    
    \item $\sup_{\{s':0< \Vert s'-s\Vert_h<1/\Vert\overline s\Vert_h\}}\frac{g(s')}{g(s)}< 
2$.
\end{enumerate}
\end{claim}

\begin{proof}Suppose not, to contradiction. Then there exists $\overline s $ with $\Vert\overline s\Vert_h>0$ and $\epsilon \in (0,1)$ with the following property: for every $s$ with $\Vert s-\bar{s}\Vert_h\ge 1$, we can find $s'$ with $\Vert s'-s\Vert_h > 0$ such that either (i) $\Vert s'-s\Vert_h \geq 1/\Vert\overline s\Vert_h$ and $\frac{g(s')}{g(s)} \ge 
\epsilon$, or (ii) $0 < \Vert s'-s\Vert_h < 1/ \Vert\overline s\Vert_h$ and $\frac{g(s')}{g(s)} \ge 2$. For an arbitrary choice of $s'$ given $s$, we define $k_s := \Vert s'-s\Vert_h$. That means, for each $s$ with $\Vert s - \overline s\Vert_h\ge 1$, we have $k_s > 0$ and a signal $s'$ with $\Vert s'-s\Vert_h=k_s$ such that either (i) $\frac{g(s')}{g(s)} \ge 
\epsilon \geq \epsilon^{k_s \Vert\overline s\Vert_h}$ (because $k_s \Vert\overline s\Vert_h \geq 1$), or (ii) $\frac{g(s')}{g(s)} \ge
2 > \epsilon^{k_s \Vert\overline s\Vert_h}$ (because $\epsilon < 1$). 

We construct a sequence of signals $(s_i)_{i=1}^{\infty}$. First, take any $s_1$ such that $\Vert s_1 -\overline s\Vert_h=1$. Then, for all $i>1$, take any $s_i$ given $s_{i-1}$ as explained in the previous paragraph. Note that for all $i$, $\Vert s_i- s_{i-1}\Vert_h = k_{s_{i-1}}$, so $\Vert s_i\Vert_h=(\Vert\overline s\Vert_h + 1) + \sum_{j=1}^{i-1} k_{s_j}$.

First, suppose that $\sum_{i=1}^{\infty} k_{s_i}=\infty$, so that $\lim_{i\to \infty} \Vert s_i\Vert_h= \infty$.  It holds that for all $s_i$, $\frac{g(s_i)}{g(\overline s)} \geq \frac{g(s_1)}{g(\overline s)} \e^{(k_{s_{i-1}} + \cdots +  k_{s_1}) \Vert \overline s\Vert_h} = \frac{g(s_1)}{g(\overline s)} \e^{(\Vert s_i\Vert_h - \Vert\overline s\Vert_h -1) \Vert\overline s\Vert_h}$, which in turn implies that
\begin{equation}
(\Vert s_i\Vert_h - \Vert\overline s\Vert_h -1)  \Vert\overline s\Vert_h \log (\epsilon) + \log(g(s_1)) \leq  \log (g(s_i)).\label{eq:subexp1}
\end{equation} 
However, since $g$ is subexponential, and $\Vert s_i\Vert_h\leq \Vert s_i\Vert \Vert h\Vert$, 
there is $p > 1$ such that for all large enough $i$,
\begin{equation}
\log (g(s_i)) < -\left(\frac{\Vert s_i\Vert_h}{\Vert h\Vert} \right)^p.\label{eq:subexp2}
\end{equation} 
The left-hand side of inequality \eqref{eq:subexp1} is linear in $\Vert s_i\Vert_h$ while the right-hand side of inequality \eqref{eq:subexp2} has exponent $p>1$, so for large enough $i$ these inequalities are in contradiction.

Next, suppose instead $\lim_{i \to \infty} \Vert s_i\Vert_h<\infty$. Then there is $N$ such that for all $i \geq N$, we have $k_{s_i} < 1/\Vert\overline s\Vert_h$ and thus $\frac{g(s_{i+1})}{g(s_i)} \ge
2$. It follows that $\lim_{i \to \infty} \frac{g(s_{i})}{g(s_N)} \ge 
\lim_{i \to \infty} 2^{i-N} = \infty$. This contradicts the boundedness of $g$ (being a density, $g$ is bounded because it is uniformly continuous).
\end{proof}

\begin{proof}[Proof of \autoref{lem:SLS}]

Without loss, we only prove that $\{\omega:h\cdot \omega > c\}$ is distinguishable from $\{\omega:h\cdot \omega< c\}$. 

We use \autoref{claim:strictly subexponential} iteratively to construct a signal sequence $(s^*_i)_{i=1}^{\infty}$. Choose any $s^*_1$ with $\Vert s^*_1\Vert_h>0$, and for $i>1$, choose any $s^*_{i}$ such that $\Vert s^*_{i}- s^*_{i-1}\Vert_h\ge 1$ that satisfies (i) $\sup_{\{s':\Vert s'-s_i^*\Vert_h\ge 1/\Vert s_{i-1}^*\Vert_h\}}\frac{g(s')}{g(s^*_{i})} < 
\frac{1}{i-1}$ and (ii) \mbox{$\sup_{\{s':0< \Vert s'-s_i^*\Vert_h< 1/\Vert s_{i-1}^*\Vert_h\}}\frac{g(s')}{g(s^*_{i})} <
2$}. This construction is well-defined by \autoref{claim:strictly subexponential}, with $\lim_{i\to \infty}\Vert s^*_i\Vert_h=\infty$.

As noted after \autoref{def:distinguishability}, it is sufficient to prove that any $\bar \omega \in \{\omega : h\cdot \omega > c\}$ is distinguishable from $\{\omega : h\cdot \omega < c\}$.\footnote{We note that this uses the assumption of countable states.}  So take any such $\bar \omega$ and $\mu$ with $\mu(\bar \omega)>0$.
Define $\overline s_i := s_i^* + \overline{\omega}$. 
It follows that for all $i$,
\begin{equation}\label{eq:near}
    \Vert\omega - \overline{\omega}\Vert_h < 0 \implies \frac{f(\overline s_i | \omega)}{f(\overline s_i | \overline{\omega})} = \frac{g(\overline s_i - \omega)}{g(\overline s_i - \overline{\omega})} = \frac{g(s_i^* + (\overline{\omega} - \omega))}{g(s_i^*)}  < 
2 ,
\end{equation}
and
\begin{equation}\label{eq:far}
    \Vert\omega - \overline{\omega}\Vert_h \le -\frac{1}{\Vert s^*_{i-1} \Vert_h} \implies  \frac{f(\overline s_i | \omega)}{f(\overline s_i | \overline{\omega})} = \frac{g(s_i^* + (\overline{\omega} - \omega))}{g(s_i^*)}  <
\frac{1}{i-1},
\end{equation}
and thus,
\begin{equation}
\label{eq:middle}
\begin{split}
& \frac{\mu(\{\omega:h\cdot \omega <c\} | \overline{s}_i)}{\mu(\overline{\omega} | \overline{s}_i)} \leq
\frac{\sum_{\Vert\omega -\overline{\omega}\Vert_h<0}\mu(\omega) f(\overline{s}_i | \omega)}{ \mu(\overline{\omega})f(\overline{s}_i | \overline{\omega})}\\[5pt]
& < \frac{1}{i-1} \frac{\sum_{\Vert\omega - \overline{\omega}\Vert_h\le - 1/\Vert s_{i-1}^*\Vert_h}\mu(\omega)}{\mu(\overline{\omega})} +2\frac{\sum_{-1/\Vert s_{i-1}^* \Vert_h < \Vert\omega -\overline{\omega}\Vert_h < 0}\mu(\omega)}{\mu(\overline{\omega})}.
\end{split}
\end{equation}
The last expression can be taken arbitrarily small because $\Vert s^*_{i-1} \Vert_h \to \infty$ as $i \to \infty$. 

It remains only to show that the above argument holds for a positive measure of signals rather than just a single $\bar s_i$. Since $g$ is uniformly continuous, there is a 
neighborhood of $\bar s_i$, say $\overline S_i$, over which \eqref{eq:near} and \eqref{eq:far} hold with slightly relaxed bounds; for instance, the bounds can be relaxed to $4$ and ${2}/({i-1})$, respectively. This establishes the analog of inequality \eqref{eq:middle} for all signals in $\overline S_i$ with the relaxed bounds. Since we have assumed $g(\cdot)>0$, each $\overline S_i$ has positive measure, so we conclude $\bar \omega$ is distinguishable from $\{\omega:h\cdot \omega< c\}$.
\end{proof}

\section{Other Material}\label{app:other}

\begin{proof}[Proof of \autoref{lem:DUB-Distinguishability}]
As noted before the lemma, $\Omega'$ is distinguishable from $\Omega''$ if and only if each $\omega'\in \Omega'$ is distinguishable from $\Omega''$. So fix any $\omega'\in \Omega'$.

We first prove that if the lemma's condition holds, then $\omega'$ is distinguishable from $\Omega''$. Take any probability measure
$\mu\in \Delta(\{\omega'\} \cup  \Omega'')$ 
such that $\mu(\omega')>0$. By assumption, for any $\varepsilon>0$ there exists a positive-probability set of signals $S'$ such that $\frac{f(s|\omega'')}{f(s|\omega')}<\varepsilon,\forall \omega'' \in \Omega'',\forall s\in S'$.
It follows that for all $s\in S'$,
$$\mu(\omega'|s)=\frac{f(s|\omega')\mu(\omega')}{\sum_{\tilde\omega \in \{\omega'\} \cup \Omega'' }f(s|\tilde\omega)\mu(\tilde\omega)}=\frac{\mu(\omega')}{\mu(\omega')+\sum_{\tilde\omega \in \Omega''}\frac{f(s|\tilde\omega)}{f(s|\omega')}\mu(\tilde\omega)}>\frac{\mu(\omega')}{\mu(\omega')+\varepsilon}.$$
Since for any $\varepsilon>0$ we can find a positive-probability set of signals $S'$ satisfying the above inequality, we conclude that 
for any $\epsilon>0$, $\Pr_\mu(s:\mu_s(\Omega')>1-\epsilon)>0$.

We next prove that if $\omega'$ is distinguishable from $\Omega''$, and $\Omega''$ is finite, then the lemma's condition holds. Consider any 
$\mu$ uniformly distributed over $\{\omega'\} \cup \Omega''$. 
The distinguishability of $\omega'$ from $\Omega''$ 
implies that for every $\epsilon > 0$ there is a positive-probability set of signals $S'$ such that $\forall s\in S'$ we have
$\frac{\sum_{\tilde\omega \in \Omega''} f(s | \tilde\omega)}{f(s | \omega')} < \epsilon$, and so $\frac{f( s | \tilde\omega)}{f(s | \omega')} < \epsilon$ for every $\tilde\omega \in \Omega''$. 
\end{proof}

\begin{remark}
    \label{rem:diffusionutility}
{\citepos{LS15} definition of diffusion utility is tailored to their binary-binary model. In general, we can define it as the highest utility an agent can obtain from any Bayes-plausible distribution of beliefs that is supported on the set of feasible posteriors (i.e., those available under the given information structure and the prior); call the corresponding signal structure the expert signal structure. 

Let us now argue that diffusion utility is lower than cascade utility. Notice that diffusion utility must be lower than first drawing a posterior from an arbitrary Bayes-plausible distribution of stationary beliefs and then drawing a signal from the expert signal structure, because this ``combined'' signal structure is Blackwell more informative than just the expert signal structure. But in the combined structure, the expert signal has no value by definition of stationary beliefs, and so the combined signal structure provides a utility equal to that from the (arbitrary Bayes-plausible)  distribution of stationary belief distributions.}    


Diffusion utility and cascade utility coincide with two states and two actions, as noted by \citepos{LS15}. But adding even one action can break this coincidence, e.g., if the third action is a ``safe'' action---one that is optimal only for some interval of interior beliefs---that shrinks the set of stationary beliefs. For a starker example, recall \autoref{eg:responsive} with $\Omega=\{0,1\}$, $A=[0,1]$, and $u(a,\omega)=-(a-\omega)^2$. Any nontrivial information structure leads to learning, with the stationary beliefs being just $0$ and $1$. So the cascade utility is the full-information utility of $0$, whereas diffusion utility will be strictly lower absent unbounded beliefs.
\end{remark}

\newpage

\renewcommand*\thesection{SA.\arabic{section}}
\setcounter{section}{0}
\renewcommand*{\theHsection}{chX.\the\value{section}}

\renewcommand*\theequation{SA.\arabic{equation}}
\setcounter{equation}{0}
\renewcommand{\theHequation}{Supplement.\theequation}

\renewcommand*\thefigure{SA.\arabic{figure}}
\setcounter{figure}{0}
\renewcommand{\theHfigure}{Supplement.\thefigure}

\renewcommand{\theexample}{SA.\arabic{example}}
\renewcommand{\theHexample}{Supplement.\theexample}
\setcounter{example}{0}

\renewcommand{\thelemma}{SA.\arabic{lemma}}
\renewcommand{\theHlemma}{Supplement.\thelemma}
\setcounter{lemma}{0}

\renewcommand{\theproposition}{SA.\arabic{proposition}}
\setcounter{proposition}{0}
\renewcommand{\theHproposition}{Supplement.\theproposition}

\renewcommand{\theconjecture}{SA.\arabic{conjecture}}
\setcounter{conjecture}{0}
\renewcommand{\theHconjecture}{Supplement.\theconjecture}

\renewcommand{\theclaim}{SA.\arabic{claim}}
\setcounter{claim}{0}
\renewcommand{\theHclaim}{Supplement.\theclaim}

\noindent {\LARGE \textbf{Supplementary Appendices
}}

\section{Belief Convergence}
\label{sec:belief convergence}

This section elaborates on \autoref{rmk:belief convergence}. Our discussion in this section focuses on deterministic networks.

One may expect the social belief to be eventually close to the stationary set with high probability: after all, when an agent's social belief is not close to the stationary set, her private information gives her a welfare improvement bounded away from zero; expanding observations should propagate these improvements, 
which implies (since utility is bounded) that they must eventually vanish. However, the following is a counterexample.\footnote{Absent expanding observations, there are trivial counterexamples using the empty network.}

\begin{example}\label{exp:belief convergence}
    Consider binary states with a uniform prior, binary signals with symmetric precision (less than 1), and binary actions with simple utility. The network is as follows: agents 1 and 2 observe no one; for odd $n\ge 3$, agent $n$ observes agent $n-2$; for even $n> 3$, agent $n$ observes agent $n-1$ and agent 2. So there is expanding observations. In this network, the odd agents form an immediate-predecessor network and there is an equilibrium where a cascade along this subsequence starts from agent 3.

    Now consider even agents. Consider the positive-probability event in which agents 1 and 2 take different actions. An even agent $n> 3$ observes agents $n-1$ and 2, which, given the equilibrium behavior of odd agents, is equivalent to observing agents 1 and 2. So the social belief of every even agent $n> 3$ equals the prior, which is bounded away from the stationary set.\footnote{The example illustrates that with positive probability social beliefs may not eventually converge to the set of stationary beliefs. But the point also holds for posterior beliefs, not just social beliefs. For simplicity, consider the same example but with an additional signal that is uninformative. Call the two actions $a$ and $b$. Consider an equilibrium in which the first agent plays $a$ upon receiving the uninformative signal, while the second agent plays $b$ upon receiving the uninformative signal. Then, in the event that the first agent plays $b$ and the second agent plays $a$, the path of social beliefs for agents $n\geq 3$ is identical to the example above: odd agents are in a cascade, while even agents' social belief is just the prior. With positive probability, an even agent will now receive an uninformative signal, whereafter her posterior belief lies outside the stationary set.} 
\end{example}


The 
``problem'' in \autoref{exp:belief convergence} is that even though each of the even agents ($n>3$) is getting a welfare improvement bounded away from zero, these improvements are not passed on to any future agents, and all future even agents continue to have social beliefs bounded away from the stationary set. 
In other words, expanding observations is not enough to validate the intuition described before the example. The following proposition identifies a reasonable condition on the network that is sufficient.


\begin{proposition}
\label{prop:beliefconvergence}
    Assume there exist finitely many subsequences of agents $\{n_{k,j}\}_{k=1}^{N_j}$ ($j=1,\ldots,J<\infty$, $1\le N_j\le\infty$) such that agent $n_{k,j}$ observes $n_{k-1,j}$, and every agent in society is in at least one of the subsequences. Then, for all $\epsilon>0$, $\lim_{n\to \infty} \phi_n(\mu_n\in S^\epsilon)=1$.
\end{proposition}

The proposition's assumption encompasses canonical examples like the complete network and the $k$-immediate-predecessor networks (i.e., every agent observes the last $k$ agents) for any $k\geq 1$. But it rules out any network in which infinite number of agents are not observed by any of their successors, which explains why it does not apply to \autoref{exp:belief convergence}.

\begin{proof}[Proof of \autoref{prop:beliefconvergence}]
    Along any subsequence $j$, $u(\phi_{n_{k,j}})\ge u(\phi_{n_{k-1,j}})+I(\phi_{n_{k-1,j}})$ by the improvement principle, given that $n_{k-1,j}$ is observable to $n_{k,j}$. It follows that $\sum_{k=1}^{N_j} I(\phi_{n_{k,j}})\le 2\bar{u}$. Hence, society's total improvement is bounded: $\sum_n I(\phi_n)\le 2\bar{u}J$. 

    Now fix any $\e,\delta>0$. Consider $V_{\delta/2}$ defined in \autoref{lemma: compact of V}. The lemma established that $V_{\delta/2}$ is compact and $\phi(\mu\notin V_{\delta/2})<\delta/2,\forall \phi\in \Phi^{BP}$. Since $S^\e$ is open, $K:=(S^\e)^c\cap V_{\delta/2}$ is compact. Next we argue $\P(\mu_n\in K\ i.o.)=0$. Suppose, to contradiction, $\P(\mu_n\in K\ i.o.)>0$. Then $\sum_n \P(\mu_n\in K)=\infty$ by the Borel-Cantelli lemma. Since $K$ is compact and $I(\cdot)>0$ on $K$, $I(\cdot)$ achieves its minimum in $K$ at some $\underline{\mu}\in K$ with $I(\underline{\mu})>0$. So the total improvement is $\sum_n I(\phi_n)\ge I(\underline{\mu})\sum_n \P(\mu_n\in K)=\infty$, which contradicts $\sum_n I(\phi_n)\le 2\bar{u}J$.

    Observe that $\P(\mu_n\in K\ i.o.)=0$ implies $\phi_n(\mu_n\in K)<\delta/2$ for all large $n$. Therefore, for all large $n$, $\phi_n(\mu_n\in (S^\e)^c)\le \phi_n(\mu_n\in K)+\phi_n(\mu_n\notin V_{\delta/2})<\delta$. We conclude that for all $\epsilon>0$, $\lim_{n\to \infty} \phi_n(\mu_n\in S^\epsilon)=1$.
\end{proof}

\begin{remark}
If $\Delta \Omega$ is compact (e.g.,  $\Omega$ itself is compact), we can replace $V_{\delta/2}$ in the proof with $\Delta \Omega$, so that $K=(S^\epsilon)^c$. Then the argument in the proof's second paragraph shows that $\P(\mu_n\in (S^\e)^c\  i.o.)=0$, i.e., the social belief converges to the stationary set almost surely rather than only in probability.
\end{remark}

\section{$\epsilon$-Excludability}
\label{sec:eps-excludability}

This section elaborates on \autoref{rem:eps-excludability}. Say that for any $\epsilon\in (0,1/2)$ a set of states $\Omega'$ is \emph{$\epsilon$-distinguishable} from $\Omega''$ if for any $\mu\in \Delta (\Omega'\cup\Omega'')$ with $\mu(\Omega')>\epsilon$, there is a positive-measure set of signals $S'$ such that $\mu(\Omega'|s)>1-\e$ for all $s\in S'$. A utility function and an information structure jointly satisfy \emph{$\e$-excludability} if $\Omega_{a_1,a_2}$ and $\Omega_{a_2,a_1}$ are $\e$-distinguishable from each other, for any pair of actions $a_1,a_2$. Note that $\epsilon$-excludability implies $\epsilon'$-excludability for all $\epsilon'>\epsilon$, and excludability is equivalent to $\epsilon$-excludability for all $\epsilon>0$.

\begin{proposition}
\label{cor:eps-distinguishability}
Let $\Omega$ be finite. For all $\epsilon \in (0,1/2)$, $\e$-excludability implies that in any equilibrium $\sigma$, $\lim \inf_n \E_{\sigma,\mu_0} [u_n]\ge u^*(\mu_0)-2\bar{u} \frac{\epsilon}{1-\e}|\Omega|$.       
\end{proposition}

Before proving \autoref{cor:eps-distinguishability}, we give an example illustrating the result's use.

\begin{example}
\label{eg:laplace}
There are three states, $\omega\in \{1,2,3\}$, SCD preferences, and Laplace information:
$$f(s|\omega)=\frac{1}{2b}\exp\left(-\frac{|s-\omega|}{b}\right),$$
where $b>0$ is a scale parameter; a smaller $b$ corresponds to more precise information.

It is straightforward to verify that no two states can be distinguished from each other.\footnote{For any pair of states $\omega \neq \omega'$, and any signal $s$, the likelihood ratio $f(s|\omega)/f(s|\omega') \leq \exp(2/b)$. 
} Therefore, not every stationary belief has adequate knowledge (so long as preferences are nontrivial), and by \autoref{lem:learning and stationary beliefs} there is inadequate learning.

Nonetheless, we claim that $\e$-excludability holds for any $\e$ such that $\e>\frac{1}{1+\exp(\frac{1}{2b})}$.  To see this, observe that since the information structure has MLRP and preferences satisfy SCD, we can focus on $\epsilon$-distinguishing state $3$ from $2$ (or, equally, $2$ from $1$).\footnote{By MLRP, only arbitrarily large signals can distinguish a state from a lower state, and for large $s$ the likelihood ratio $f(s|3)/f(s|2)<f(s|3)/f(s|1)$, so considering adjacent states is sufficient for $\e$-excludability.} When $\e>\frac{1}{1+\exp(\frac{1}{2b})}$, we have $\frac{\e}{1-\e}\exp(1/b)>\frac{1-\e}{\e}$, so there exist signals that move the prior $(0,1-\e,\e)$ to a posterior of at least $1-\epsilon$ on state $3$, which implies $\e$-distinguishability of state $3$ from $2$.

\autoref{cor:eps-distinguishability} implies that in any equilibrium, $\lim \inf \E_{\sigma,\mu_0} [u_n]\ge u^*(\mu_0)-6\bar{u}\exp(-\frac{1}{2b})$. This quantitative welfare bound yields, in particular, convergence to the full-information utility $u^*(\mu_0)$ as $b\to 0$.
\end{example}

\begin{proof}[Proof of \autoref{cor:eps-distinguishability}]
Take any stationary belief $\mu$, and let $a$ be an optimal action at belief $\mu$. For each state $\omega$, take any $a_{\omega} \in c(\omega)$, and consider $\mu_\omega(\cdot):=\mu(\cdot|\{\omega\}\cup \Omega_{a,a_{\omega}})$. If $\mu_\omega(\omega)\le \e$, then $\mu(\omega)\le \e$, so $(u(a_{\omega},\omega)-u(a,\omega))\mu(\omega)\le 2\bar{u}\e$. 

Consider the other case of $\mu_\omega(\omega)> \e$. For any $s \in S$, because $u(a,\omega') - u(a_\omega,\omega') \leq 0$ for each $\omega' \notin \Omega_{a, a_{\omega}}$, and $\mu$ is stationary, 
$$\sum_{\omega'\in \{\omega\}\cup \Omega_{a,a_{\omega}}} (u(a,\omega')-u(a_\omega,\omega'))\mu(\omega'|s) \geq \sum_{\omega'\in\Omega} (u(a,\omega')-u(a_{\omega},\omega'))\mu(\omega'|s) \geq 0.$$
Then,
\begin{eqnarray*}
(u(a_{\omega},\omega)-u(a,\omega))\mu_\omega(\omega|s) & \le& \sum_{\omega'\in \Omega_{a,a_{\omega}}} (u(a,\omega')-u(a_{\omega},\omega'))\mu_\omega(\omega'|s) \\
& \leq & 2 \overline{u} \left(\sum_{\omega' \in \Omega_{a, a_{\omega}}} \mu_{\omega}(\omega' | s)\right) = 2 \overline{u} (1-\mu_{\omega}(\omega | s)).
\end{eqnarray*}
By $\e$-excludability, there exists a positive-measure set of signals $S'$ such that, for any $s \in S'$, $\mu_\omega(\omega|s)>1- \e$, which implies that $u(a_{\omega},\omega)-u(a,\omega) \leq 2\overline{u}\frac{\e}{1-\e}$.

In either case ($\mu_{\omega}(\omega) \leq \e$ or $\mu_{\omega}(\omega) > \e$), we have  $(u(a_{\omega},\omega)-u(a,\omega))\mu(\omega)\le 2\bar{u}\frac{\e}{1-\e}$. Since $\Omega$ is finite,
$$\sum_{\omega \in \Omega}(u(a_{\omega},\omega)-u(a,\omega))\mu(\omega)\le 2\bar{u}\frac{\e}{1-\e}|\Omega|.$$
Namely, the utility gap $u^*(\mu)-u(\mu)\le 2\bar{u}\frac{\e}{1-\e}|\Omega|$, for any stationary belief $\mu$.

Finally, for any $\phi\in \Phi^S$,
$$u^*(\mu_0)-u(\phi)=\E_\phi[u^*(\mu)-u(\mu)]\le 2\bar{u}\frac{\e}{1-\e}|\Omega|.$$
By taking infimum of $u(\phi)$ across $\phi \in \Phi^{S}$, we obtain $u_*(\mu_0)\ge u^*(\mu_0)-2\bar{u}\frac{\e}{1-\e}|\Omega|$, and subsequently by invoking \autoref{thm:utility diffusion},
we conclude that in any equilibrium $\sigma$, $\lim \inf_n \E_{\sigma,\mu_0} [u_n]\ge u^*(\mu_0)-2\bar{u}\frac{\e}{1-\e}|\Omega|$.
\end{proof}

\section{Details on \hyperref[eg:DUBnoSCD]{Example} \ref{eg:DUBnoSCD}}\label{sec:eg2}
For \autoref{eg:DUBnoSCD}, we show here how to construct a full-support prior such that the posterior probability is uniformly bounded away from $1$ across signals and states. Take any prior $\mu$ such that for some $c>0$, $\min\left\{\frac{\mu(n-1)}{\mu(n)},\frac{\mu(n+1)}{\mu(n)}\right\}>c$ for all $n$ (e.g., a double-sided geometric distribution). Denoting the posterior after signal $s$ by $\mu_s$, the posterior likelihood ratio satisfies $$\frac{\mu_s(\{n-1,n+1\})}{\mu_s(n)}=\frac{f(s|n-1)}{f(s|n)}\frac{\mu(n-1)}{\mu(n)}+\frac{f(s|n+1)}{f(s|n)}\frac{\mu(n+1)}{\mu(n)}>c \left(\frac{f(s|n-1)}{f(s|n)}+\frac{f(s|n+1)}{f(s|n)}\right).$$
As the last expression is the sum of a strictly positive decreasing function of $s$ and a strictly positive increasing function of $s$, it is bounded away from $0$ in $s$. The bound is independent of $n$ because normal information is a location-shift family of distributions. Therefore, the posterior likelihood ratio is uniformly bounded away from $0$, and hence, the posterior $\mu_s(n)$ is uniformly bounded away from $1$.

\section{Learning at a Fixed Prior}
\label{app:fixedprior}

For tractability, our discussion in this section assumes a complete network. 

\paragraph{The issue.} The definition of adequate learning we adopted in \autoref{sec:model} requires that there is learning for all priors. In general, one may be interested in whether there is adequate learning at some given (full-support) prior.\footnote{To be complete:  there is \emph{adequate learning at prior $\mu_0$} if for every equilibrium strategy profile $\sigma$, 
$\E_{\sigma, \mu_0} [u_n] \to u^*(\mu_0)$. Adequate (or inadequate) learning at a prior for a choice set is defined analogously.} Of course, our sufficient conditions---e.g., \autoref{thm:exclude}'s excludability---remain sufficient, but fixing a prior raises the question of whether the conditions are necessary.  With only two states, the distinction between some prior and all priors is immaterial: if adequate learning fails at any prior, then the only adequate-knowledge beliefs are those with certainty on some state, and there is an open ball of stationary beliefs around certainty on one of the states; hence, given any full-support prior, there cannot be a belief path that converges to certainty on that state, implying a failure of adequate learning at all full-support priors.

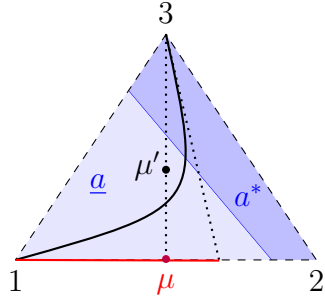
\begin{figure}[h] 
\centering 
		\begin{tikzpicture}
		\coordinate (A) at (0,0); \coordinate (B) at (4,0); \coordinate (C) at (2,3); \coordinate (D) at (2.8,1) ; \coordinate (E) at (2,0); 
			\coordinate (prior) at (2,1.2);
			\coordinate (midpriorB) at (2.6,.45);); 
			\coordinate (muend) at ($(C)!1.18!(midpriorB)$;) ;
			\coordinate (F) at ($(A)!1.43!(prior)$);); 
			\coordinate (a32) at (3.4,0) ;
			\coordinate (a31) at (1.5,2.25) ;
			\coordinate (aunder) at (1.5,1.0);
			\coordinate (astar) at (3.1,1.3);
			\coordinate (prior12) at ($(C)!1.66!(prior)$;); 			
			\coordinate (belief12) at (1.1,0);

		\fill[blue!25] (a31) -- (a32) -- (B) -- (C);
		\draw[color=blue] (a31) -- (a32);
		\node[label={[blue] below:$a^*$}] at (astar) {};

		\fill[blue!10] (B) -- (A) -- (a31) -- (a32);	
		\node[label={[blue] left:$\underline{a}$}] at (aunder) {};

		\draw[dashed] (A) node[below] {$1$} -- (B) node[below] {$2$} -- (C) node[above] {$3$} -- (A) ;
		\draw[thick,dotted] (C) -- (muend);	
		\draw[thick,dotted] (C) -- (prior12);	

		\draw[thick] (A) .. controls (2.5,.7) .. (C);

		\node[circle,fill=black,inner sep=0pt,minimum size=3pt,label={[xshift=-7, yshift=-10]:{$\mu'$}}] (a) at (prior) {};

		\draw[red,thick] (muend) -- (A);
		\node[circle,fill=purple,inner sep=0pt,minimum size=3pt,label={[red] below:{$\mu$}}] at (prior12) {};

		\end{tikzpicture}
\caption{Preference regions among the actions $\underline a$ and $a^*$ shaded in blues. 
Under belief $\mu'$, posteriors are given by the black curve, while under belief $\mu$, posteriors are given by the red line. }
\label{fig:DUBnecessity} 
\end{figure}

However, with multiple states, a failure of adequate learning at some prior does not imply an open ball of stationary beliefs around any adequate-knowledge belief. To illustrate, consider \autoref{fig:DUBnecessity}. Action $a^*$ is optimal at states $2$ and $3$ while $\underline a$ is optimal at state $1$. Adequate learning fails when the prior is $\mu$ because $\mu$, which has support $\{1,2\}$, is stationary but has inadequate knowledge.\footnote{In this example, preferences have SCD. But, consistent with \autoref{thm:main}, DUB is violated because state $2$ is not distinguishable from state $1$ or state $3$.} Yet there is no open ball of stationary beliefs: no full-support belief is stationary because the optimal actions are distinct at the extreme states $1$ and $3$, and the extreme states are distinguishable from their complements.  This raises the possibility that there is learning at some---or even all---full-support priors, with on-path sequences of beliefs (which necessarily have full support at every finite time) converging almost surely to adequate-knowledge beliefs without ever hitting any stationary belief (all of which have non-full-support). 

\paragraph{Some partial analysis.}
While we are unable to characterize learning at a fixed prior in general, we provide some partial analysis below that we hope will be useful for future research. We focus on obtaining an analog of \autoref{thm:exclude} for a fixed prior.

First, we provide a lemma (\autoref{lem:distinguishability and history}) that connects the existence of certain on-path histories to distinguishability. Second, we conjecture a result (\autoref{conjecture}) and show in \autoref{exclude-fixedprior} that, if the result is true, it combines with the lemma to deliver a fixed-prior analog of \autoref{thm:exclude}. Third, we show that the conjecture is true in a class of problems (\autoref{prop:noDUBMLRP}).

\begin{lemma}
\label{lem:distinguishability and history}
Take an aribtrary $\omega^*\in \Omega$ and set of states $\Omega' \subseteq \Omega \backslash \{\omega^*\}$.  State $\omega^*$ is distinguishable from $\Omega'$ if there exist an equilibrium under a full-support prior and a history of actions $h^\infty$ 
such that $\P(h^\infty|\omega^*)=0$,
and 
$\P(h^\infty| \omega)$
is bounded away from 0 across $\omega \in \Omega'$.
\end{lemma}

The lemma ties the asymptotic probabilities of on-path histories to the information structure of an individual agent. The formal proof of the lemma is provided at the end of this appendix, but to see the intuition, suppose that the relevant $h^\infty$ in the hypothesis of the lemma is some eventual herd on some action $a\in A$, i.e., $h^\infty=\{h^m,a,a,a,,...\}$ with $h^m$ some finite subhistory. If $h^\infty$ has $0$ probability in state $\omega^*$, it must be because in an infinite number of periods, agents have positive probability of obtaining signals which overturn the herd on $a$, i.e., result in them taking some other action than $a$. However, the probability of this history is positive for states in $\Omega'$. This means that the probability of signals that overturn the herd must vanish over time at a fast enough rate in $\omega^*$, but either not vanish or vanish at a slow enough rate in each $\omega \in \Omega'$. In particular, there must exist overturning signals whose probability gets arbitrarily large in state $\omega^*$ relative to those in every $\omega \in \Omega'$, which means $\omega^*$ is distinguishable from $\Omega'$.

\begin{conjecture}
\label{conjecture}
Take any $a_1,a_2 \in A$, any full-support prior, and any equilibrium. 
If there is adequate learning at that prior, choice set $\{a_1,a_2\}$, and equilibrium,\footnote{That is, under the given prior $\mu_0$ and equilibrium $\sigma$,  
$\E_{\sigma, \mu_0} [u_n] \to u^*(\mu_0)$.} then
\begin{align}
\exists h^\infty \text{ and } \epsilon>0: \ \P(h^\infty|\omega)>\epsilon,  \forall \omega \in \Omega_{a_1,a_2}.\label{eqn: extracond}
\end{align}
\end{conjecture}

The conjecture says that given any full-support prior, any binary choice set $\{a_1,a_2\}$, and any equilibrium in which there is adequate learning, we can find a single history that occurs with probability bounded away from $0$ in all states in which $a_1$ is strictly preferred (and analogously, a different history for the states in which $a_2$ is strictly preferred). To appreciate the conjecture, let us focus for discussion on the case of finite states, nontrivial information, and nontrivial preferences. First note that if $\Omega_{a_1,a_2}$ is a singleton---as is the case with binary states---then it is straightforward that there is such a history, as there is a herd almost surely and every herd begins at some finite time. When $\Omega_{a_1,a_2}$ is not a singleton, given adequate learning, the same logic shows that for each state in $\Omega_{a_1,a_2}$, there is a history that has positive probability in that state, namely one with a herd on $a_1$. But \autoref{conjecture} demands more: a single history that has 
positive probability in all states in $\Omega_{a_1,a_2}$. Nonetheless, the conjecture seems intuitive: (up to tie-breaking issues) it would be surprising for every infinite history that has positive probability in some $\omega \in \Omega_{a_1,a_2}$ to have zero probability in some other $\omega' \in \Omega_{a_1,a_2}$, given that agents' have the same ordinal preferences over the binary actions in both $\omega$ and $\omega'$. For instance, consider a fully-informative information structure and any nontrivial preferences. Clearly, given any choice set $\{a_1,a_2\}$, there are only two possible histories: either $a_1$ in every period or $a_2$ in every period. The former has probability $1$ in each $\omega\in\Omega_{a_1,a_2}$, and the latter has probability $1$ in each $\omega\in\Omega_{a_2,a_1}$ and so \autoref{conjecture} holds. Even though individuals' private information distinguishes states perfectly, the public history does not. 

\begin{proposition}
\label{exclude-fixedprior}
If \autoref{conjecture} is true, then not only does excludability imply adequate learning at every prior for every choice set, but moreover, if excludability fails, 
then there exists a choice set with inadequate learning at every full-support prior.
\end{proposition}

\begin{proof}
That excludability implies adequate learning at every prior for every choice set is implied by \autoref{thm:exclude}, with no need to invoke \autoref{conjecture}. So we only prove the second portion of the proposition, doing so by contraposition.

To that end, assume that for every choice set, there is some full-support prior at which there is adequate learning in some equilibrium. 
For every binary choice set $\{a_1, a_2\}$, \autoref{conjecture} implies the existence of a history $h^\infty$ satisfying \eqref{eqn: extracond} at the full-support prior and equilibrium at which there is adequate learning. Since there is adequate learning, eventually all agents must be taking $a_1$ in $h^\infty$, which implies that 
$\P(h^\infty|\omega^*)=0$ for each $\omega^* \in \Omega_{a_2,a_1}$. Then, taking $\Omega'=\Omega_{a_1,a_2}$ in \autoref{lem:distinguishability and history} yields that $\omega^*$ is distinguishable from $\Omega_{a_1,a_2}$. Since $a_1, a_2$ and $\omega^* \in \Omega_{a_1, a_2}$ are arbitrary,  there is excludability.
\end{proof}

We have not been able to establish \autoref{conjecture} in general. However, we are able to establish it when preferences satisfy SCD and the information structure satisfies the strict MLRP (assuming, only for convenience, that the state space is discrete):

\begin{claim}
\label{prop:noDUBMLRP}
Assume $\Omega\subseteq\mathbb{Z}$. If preferences satisfy SCD and the information structure satisfies the strict MLRP, then \autoref{conjecture} is true.
\end{claim}

The proof is at the end of this appendix. Combining \autoref{prop:noDUBMLRP} and \autoref{exclude-fixedprior},  we see that under a complete network, the signal structure and preferences in \autoref{fig:DUBnecessity} entail inadequate learning at every full-support prior, such as $\mu'$ in the figure. Note that the figure's signal structure satisfies strict MLRP because the black curve in \autoref{fig:DUBnecessity} is concave vis-\`a-vis the $1$--$3$ edge and approaches the $1$ and $3$ vertices.

\subsection*{Omitted Proofs}

\begin{proof} [Proof of \autoref{lem:distinguishability and history}]
Suppose not. Then there exist a belief \mbox{$\mu \in \Delta(\Omega' \cup \{\omega^*\})$} with $\mu(\omega^*) > 0$
and a small $\epsilon > 0$ such that for almost every signal $s$, the posterior $\mu(\omega^*| s) \le 1-\epsilon$.
By taking the conditional distribution of $\mu$ on $\Omega'$, call it $\tilde{\mu}$, and $z := \frac{\epsilon\mu(\omega^*)}{1-\mu(\omega^*)}  \in (0,1)$, we obtain for almost every $s$, 
\begin{equation}
\int_{\Omega'} f(s|\omega)
\dif \tilde{\mu}(\omega) \geq z f(s | \omega^*).
\label{eq:not_distinguishable}
\end{equation}

Suppose there exist an equilibrium $\sigma$ under a full support prior and history $h^\infty$ such that $\P(h^\infty|\omega^*)=0$, and 
$\P(h^\infty| \omega)$ is bounded away from 0 across $\omega \in \Omega'$. Let $a^n$ be the action taken by agent $n$ along $h^\infty$ and $A^{-n}:=A\backslash \{a^n\}$. Let $\P(a^n | h^n, \omega) :=\int_S \sigma(a^n | s, h^n)f(s|\omega) \dif s$
be the probability that agent $n$ plays action $a^n$ when the state is $\omega$ and the sub-history is $h^n$.  It holds that:
\allowdisplaybreaks
\begin{align}
\sum_{n=1}^{\infty} \log(1- z \P(A^{-n} | h^n, \omega^*)) & \geq \sum_{n=1}^{\infty} \log\left(1 - \int_{\Omega'}  \P(A^{-n} | h^n, \omega)\dif \tilde{\mu}(\omega)\right) \qquad \text{(using \eqref{eq:not_distinguishable})} \notag\\[5pt]
 & = \sum_{n=1}^{\infty} \log\left(\int_{\Omega'}  \P(a^n | h^n, \omega)\dif \tilde{\mu}(\omega) \right) \notag\\[5pt]
& \geq \sum_{n=1}^{\infty} \int_{\Omega'}  \log(\P(a^n | h^n, \omega))\dif \tilde{\mu}(\omega) \qquad \text{(by Jensen's inequality)} \notag\\[5pt]
& = \int_{\Omega'}  \sum_{n=1}^{\infty}   \log(\P(a^n | h^n, \omega))\dif \tilde{\mu}(\omega) \qquad \text{(by Tonelli's theorem)} \notag\\[5pt]
& =\int_{\Omega'}  \log\left(\prod_{n=1}^{\infty} \P(a^n | h^n, \omega)\right)\dif \tilde{\mu}(\omega) \notag\\[5pt]
& > -\infty \qquad \text{(as $\log\P(h^\infty|\omega)$ is bounded across $\omega\in {\Omega'}$)}. \label{eqn:finite_log_sum} 
\end{align}

Below we will invoke the fact that for arbitrary sequences $(S_n)$ and $(T_n)$ and constant $c>0$, if $\lim_{n\to \infty} \frac{S_n}{T_n}=c>0$ and $\sum_{n} S_n<\infty$, then $\sum_{n} T_n<\infty$.\footnote{For any $c'<c$ there exists $N$ such that for all $n > N$, $S_n/T_n \geq c'$, or $T_n\leq S_n/c'$. So $\sum_{n}T_n \leq \sum_{n \leq N} T_n + \sum_{n>N} S_n/c'<\infty$.} Let $S_n=-\log(1-z\P(A^{-n}|h^n,\omega^*))$ and $T_n=-\log(1-\P(A^{-n}|h^n,\omega^*))$. Note that $\lim_{n\to \infty} \frac{S_n}{T_n}=z\in (0,1)$ because $\lim_{x \to 0} \frac{\log(1-z x)}{\log(1-x)} = z$  and \eqref{eqn:finite_log_sum} implies $\lim_{n \to \infty} \P(A^{-n} | h^n, \omega^*) = 0$. The aforementioned mathematical fact implies that 
$$\sum_{n=1}^{\infty} \log(1-\P(A^{-n}|h^n,\omega^*)) >-\infty.$$
As $\P(a^n|h^n,\omega^*)=1-\P(A^{-n}|h^n,\omega^*)$, it further follows that $\prod_{n=1}^{\infty} \P(a^n | h^n, \omega^*)>0$, which contradicts $\P(h^\infty|\omega^*)=0$. 
\end{proof}

\begin{proof}[Proof of \autoref{prop:noDUBMLRP}]
\label{proof:noDUBMLRP}
Take any information structure $f$ with strict MLRP, and a utility function $u$ that has SCD. Then, take an equilibrium $\sigma$ under a full support prior $\mu$ and a binary choice set $\{a_1, a_2\}$. Since $u$ has SCD, $\Omega_{a_1,a_2}$ and $\Omega_{a_2,a_1}$ are either an upper and lower set or the other way around. We consider the case that $a_2$ is preferred in higher states and $a_1$ is preferred in lower states. We omit an analogous proof for the other case.

We observe that $\P(a_1 | h^n, \omega) := \int_S \sigma(a_1 | h^n, s) f(s | \omega)\dif s$, the probability that agent $n$ plays action $a_1$ given any finite history $h^n$, decreases in $\omega$. First, the probability $\sigma(a_1 | h^n, s)$ decreases in $s$. The social belief $\mu(\cdot | h^n)$ has full support, so strict MLRP of the information structure implies that $\forall s < s'$, the posterior $\mu(\cdot | h^n, s')$ strictly monotone likelihood-ratio dominates $\mu(\cdot | h^n, s)$. By Theorem 2 of \citet{athey2002mcs},
$$
D(s) := \sum_\omega \left(u(a_2,\omega)-u(a_1,\omega)\right)\mu(\omega|h^n,s)
$$
is strictly single crossing in $s$, i.e., $D(s)\geq 0 \implies D(s')>0$, $\forall s'>s$. Hence, $\sigma(a_1|h^n, s)$ is decreasing in $s$. Since $f$ satisfies strict MLRP, $\P(a_1 | h^n, \omega)$ is decreasing in $\omega$.

Next, suppose there is adequate learning.
So for each state $\omega\in\Omega_{a_1,a_2}$, there is an infinite history with a herd on $a_1$, $h^\infty=(\ldots,a_1,a_1,\ldots)$
that occurs with positive probability in $\omega$. In particular, if we let $\tilde\omega=\max\Omega_{a_1,a_2}$, 
then for any finite sub-history $h^n$ of $h^{\infty}$,
$$\forall \omega\le \tilde\omega: \P(a_1 | h^n, \omega)\geq \P(a_1 | h^n, \tilde\omega)>0,$$
and since $h^\infty$ has positive probability at $\tilde\omega\in\Omega_{a_1,a_2}$, it follows that
\begin{equation}
\label{eq:posprob}
\forall \omega\le \tilde \omega: \prod_{n=1}^{\infty} \P(a_1 | h^n, \omega) \geq \prod_{n=1}^{\infty} \P(a_1 | h^n, \tilde\omega) > 0.
\end{equation}
This means $\P(h^\infty|\omega)$ is uniformly bounded away from 0 for $\{\omega:\omega\le \tilde\omega\}$. Since $\Omega_{a_1,a_2}\subseteq \{\omega:\omega\le \tilde\omega\}$, this establishes \autoref{conjecture}.  
\end{proof}

\end{document}